\documentclass[12pt]{article}
\usepackage{amssymb}
\usepackage{amsfonts}
\usepackage{youngtab}
\usepackage{amsmath,amssymb,latexsym,cite}
\usepackage{amsthm}
\usepackage{cite}
\usepackage[]{hyperref}
\input xy
\xyoption{all}

\numberwithin{equation}{section}

\begin{document}

\vspace*{0.5in}

\begin{center}

{\large\bf A proposal for nonabelian (0,2) mirrors}

\vspace{0.2in}

Wei Gu$^1$, Jirui Guo$^2$, Eric Sharpe$^1$

\begin{tabular}{cc}
{ \begin{tabular}{c}
$^{1}$ Dep't of Physics\\
Virginia Tech\\
850 West Campus Dr.\\
Blacksburg, VA  24061
\end{tabular} } &
{ \begin{tabular}{c}
$^2$ Dep't of Physics \\
Center for Field Theory \& Particle Physics \\
Fudan University \\
220 Handan Road \\
200433 Shanghai, China
\end{tabular} }
\end{tabular}

{\tt weig8@vt.edu}, {\tt jrguo@fudan.edu.cn}, {\tt ersharpe@vt.edu}

$\,$

\end{center}

In this paper we give a proposal for mirrors to
(0,2) supersymmetric gauged linear sigma models (GLSMs), 
for those (0,2) GLSMs which are
deformations of (2,2) GLSMs.
Specifically, we propose a construction of (0,2) mirrors for (0,2) GLSMs
with $E$ terms that are linear and diagonal, reducing to both the
Hori-Vafa prescription as well as a recent (2,2) nonabelian mirrors proposal
on the (2,2) locus.
For the special case of
abelian (0,2) GLSMs, two of the authors have previously proposed a
systematic construction, 
which is both simplified and generalized by the proposal
here.

\begin{flushleft}
August 2019
\end{flushleft}

\newpage

\tableofcontents

\newpage

\section{Introduction}

One of the outstanding problems in heterotic string compactifications is to
understand nonperturbative effects due to worldsheet instantons.
For type II strings and (2,2) worldsheet theories, these effects are
well-understood, and are encoded in quantum cohomology rings and Gromov-Witten
theory.  In principle, there are analogues of both for more general
heterotic theories, but there are comparatively many open questions.

For example, in a heterotic $E_8 \times E_8$ compactification 
on a Calabi-Yau threefold with a rank three bundle, the low-energy
theory contains states in the ${\bf 27}$ and ${\bf \overline{27}}$ 
representations of $E_6$, with cubic couplings appearing as spacetime
superpotential terms.  On the (2,2) locus
(the standard embedding, where the gauge bundle equals the tangent
bundle), for the case of the quintic threefold,
those couplings have the standard form 
\cite{Strominger:1985ks,Candelas:1990rm}
\begin{equation}  \label{eq:quintic-gw}
{\bf \overline{27}}^3 \: = \: 5 \: + \:
\sum_{k=1}^{\infty} n_k \frac{k^3 q^k}{1 - q^k} \: = \:
5 \: + \: 2875 \, q \: + \: 4876875 \, q^2 \: + \: \cdots,
\end{equation}
where the $n_k$ encode the Gromov-Witten invariants.
These are computed by three-point functions in the A model topological
field theory on the worldsheet.
Off the (2,2) locus, for more general gauge bundles, these couplings
have a closely analogous form: a classical contribution plus a sum
of nonperturbative contributions, without any perturbative 
loop corrections 
\cite{Wen:1985jz,Dine:1986zy,Dine:1987bq,Silverstein:1995re,Berglund:1995yu,Beasley:2003fx}.
As a result, we know that more general
heterotic versions of the Gromov-Witten invariants exist, and from
general holomorphy arguments, must be nontrivial.

In principle, off the (2,2) locus,
heterotic Yukawa couplings such as those in equation~(\ref{eq:quintic-gw})
are computed by the A/2 and B/2 models, which on the (2,2) locus
reduce to the ordinary A and B model topological field theories.
For e.g. Fano spaces, both these heterotic Gromov-Witten invariants\footnote{
For three-point functions on $S^2$, only.
More precisely, correlation function computations are understood in
(analogues of) topological field theories, but correlation function
computations in analogues of topological string theories -- with the
exception of those that reduce to topological field theory computations
-- are still unknown.
} as well
as heterotic versions of quantum cohomology rings
(here, a  quantum-corrected ring of sheaf 
cohomology groups \cite{Distler:1987ee}
of the form $H^{\bullet}(X, \wedge^{\bullet} {\cal E}^*)$,
introduced in
\cite{Katz:2004nn,Adams:2005tc,Sharpe:2006qd,Guffin:2007mp}) are known
for toric varieties \cite{McOrist:2007kp,McOrist:2008ji,Donagi:2011uz,Donagi:2011va,Closset:2015ohf},
Grassmannians $G(k,n)$ \cite{Guo:2015caf,Guo:2016suk}, and flag manifolds \cite{Guo:2018iyr},
all for the case that the gauge bundle is a deformation of the tangent
bundle.  (Cases involving more general gauge bundles are not currently
understood.)  See for example 
\cite{McOrist:2010ae,Guffin:2011mx,Melnikov:2012hk,Melnikov:2019tpl} 
for reviews.

For Calabi-Yau spaces, one can compute many correlation functions;
however, it is not yet known how to extract the 
precise analogues of Gromov-Witten
invariants from these computations, as one needs, for example, a heterotic
analogue of the Picard-Fuchs equations to get a precise mirror map and
vacuum normalization.
Similarly, more recent methods applying supersymmetric
localization \cite{Jockers:2012dk} are also not applicable.

Historically, Gromov-Witten invariants in (2,2) supersymmetric theories
were first computed using mirror symmetry, and so one might hope that
a (0,2) supersymmetric version of mirror symmetry might aid in such
developments.
This is one of the motivations to understand (0,2) mirrors
(see e.g. \cite{Blumenhagen:1996vu,Blumenhagen:1996tv} for some early work).

To date, there has been significant progress on understanding (0,2)
mirror symmetry, but many results are still limited
(and certainly
heterotic Gromov-Witten invariants for Calabi-Yau's are not yet known).  
For example,
for the case of reflexively-plain polytopes, and bundles that are
deformations of the tangent bundle, a generalization of the Batyrev
construction of ordinary Calabi-Yau mirrors exists, see
\cite{Melnikov:2010sa,Bertolini:2018qlc,Bertolini:2018usi}.

In this paper, we shall propose what is ultimately a (0,2) analogue of the
Hori-Vafa construction \cite{Hori:2000kt,Morrison:1995yh}, 
which is to say, a mirror construction
for two-dimensional gauge theories, resulting in a Landau-Ginzburg model,
which in our case will be defined for a special class of (0,2) deformations
off of the (2,2) locus.
For abelian theories, there has been nontrivial work in this area
in the past \cite{Adams:2003zy,Chen:2016tdd,Chen:2017mxp,Gu:2017nye}.
This work has included ansatzes for various special cases of
toric varieties \cite{Chen:2016tdd,Chen:2017mxp}, as well as a more
systematic proposal for abelian theories \cite{Gu:2017nye}.
The proposal in this paper will both extend such constructions to
nonabelian\footnote{
Specifically, on the (2,2) locus, this will reduce to the nonabelian
mirrors proposal described in  \cite{Gu:2018fpm,Chen:2018wep,Gu:2019zkw}.
Other proposals have appeared in the math community in e.g.
\cite{rietsch1,teleman}, as reviewed in \cite{Gu:2018fpm}[section 4.9,
appendix A].
} (0,2) GLSMs, as well as give a simpler, more straightforward,
presentation in
abelian cases than that in \cite{Gu:2017nye}.
We do not claim to have a proof of the construction, but we do show
that the proposal satisfies a number of general consistency tests consistent
with (2,2)-supersymmetric gauge theory mirrors, for
example:
\begin{itemize}
\item axial anomalies of the original theory are realized by 
classically-broken symmetries of the mirror, and can be restored by
a shift of the mirror to the theta angle,
\item quantum sheaf cohomology ring relations of the original gauge theory
are realized classically in the mirror by critical locus relations,
\item correlation functions match,
\item integrating out matter fields from the mirror returns the
one-loop effective superpotential of the original gauge theory on the
Coulomb branch,
\end{itemize}
all just as happens in (2,2)-supersymmetric gauge theory
mirrors \cite{Hori:2000kt,Gu:2018fpm}.  We also check the mirror
construction in several concrete examples.

We begin in section~\ref{sect:review-22-mirror} by reviewing
mirrors to (2,2)-supersymmetric gauge theories, both abelian
\cite{Hori:2000kt} and nonabelian \cite{Gu:2018fpm}.  We discuss the
mirror constructions themselves and expected properties of mirrors to
two-dimensional gauge theories.

In section~\ref{sect:proposal} we describe our proposal for (0,2) mirrors
for a special class of (0,2) deformations off the (2,2) locus.
As many subtleties of nonabelian mirrors have already been
extensively discussed in \cite{Gu:2018fpm,Chen:2018wep,Gu:2019zkw}, 
here we focus solely on the novel
aspects introduced by (0,2) supersymmetry.  We also discuss mirrors to
symmetries
and their anomalies, and check that the quantum sheaf cohomology ring
relations of the original theory are correctly reproduced in the
mirror by classical critical locus relations.

In section~\ref{sec:justification}, we give further general arguments checking
this proposal.  Specifically, we check that correlation functions
match, and we demonstrate that integrating-out matter fields correctly
reproduces the one-loop effective superpotential on the Coulomb branch,
exactly as happens in (2,2)-supersymmetric gauge theory mirrors.
(We do not, however, claim to have a proof.)

In section~\ref{sect:specialize}, we specialize to 
abelian theories.  In particular, the ansatz here simplifies and
generalizes the ansatz two of the authors previously discussed
in \cite{Gu:2017nye}.

In section~\ref{sect:previous} we compare to the previous systematic proposal
for (0,2) mirrors to abelian theories
by two of the authors \cite{Gu:2017nye}.  The ansatz presented here is
both more general and rather simpler, and we also argue
that when we restrict to
(0,2) deformations of the form considered in \cite{Gu:2017nye}, our
current proposal gives the same results as \cite{Gu:2017nye}.

In the next several sections, we discuss concrete examples.
We begin in section~\ref{sect:pnpm} by giving a detailed analysis of
mirrors to ${\mathbb P}^n \times {\mathbb P}^m$.  We verify correlation
functions in the original theory, construct lower-energy Landau-Ginzburg
theories in the style of (2,2) Toda duals to projective spaces, discussing
subtleties that arise in their construction, explicitly verify
correlation functions in those lower-energy theories, and also compare
to previous (0,2) mirrors for these spaces in \cite{Chen:2016tdd}.

In section~\ref{sect:hirzebruch} we perform analogous analyses for
(0,2) mirrors to Hirzebruch surfaces, constructing lower-energy theories
and comparing to results in \cite{Chen:2017mxp}.

In section~\ref{sect:grassmannian}, we discuss our first nonabelian
examples, GLSMs for (0,2) deformations of Grassmannians $G(k,N)$.
These are two-dimensional $U(k)$ gauge theory with matter in copies of the
fundamental representation.  We construct lower-energy Landau-Ginzburg
models, analogues of (2,2) Toda duals, that generalize the
Grassmannian mirrors discussed in \cite{Gu:2018fpm}, and explicitly verify that
quantum sheaf cohomology rings \cite{Guo:2015caf,Guo:2016suk} are reproduced.  
We also explicitly verify
that correlation functions are correctly reproduced in a few tractable
examples.

In section~\ref{sect:flag} we briefly discuss (0,2) deformations of
flag manifolds, generalizations of Grassmannians that are also
described by two-dimensional nonabelian gauge theories.
We verify that quantum sheaf cohomology rings \cite{Guo:2018iyr}
are reproduced.

Finally, in section~\ref{sect:hypersurfaces}, we briefly discuss
(0,2) mirrors to theories with hypersurfaces.  The rest of the paper
is concerned with mirrors to theories without a (0,2) superpotential;
in this section, we discuss how the result is modified to take into
account a (0,2) superpotential, and also discuss how the mirror ansatz
reproduces some conjectures regarding hypersurface mirrors in
\cite{McOrist:2008ji}.

\section{Review of mirrors to (2,2) supersymmetric gauge theories}
\label{sect:review-22-mirror}

In this section we shall review results of \cite{Hori:2000kt,Gu:2018fpm} on
mirror symmetry for two-dimensional (2,2) supersymmetric gauge
theories.

Briefly, in these papers, the mirror to a two-dimensional gauge
theory is given as a Landau-Ginzburg model, whose classical physics
encodes the quantum physics of the original gauge theory.

For abelian two-dimensional (2,2) supersymmetric gauge theories,
mirrors were described in \cite{Hori:2000kt}.  For a $U(1)^k$ gauge
theory with $n$ chiral superfields with charges encoded in charge
matrix $\rho_i^a$ ($a \in \{1, \cdots, k\}$, $i \in \{1, \cdots n\}$,
and Fayet-Iliopoulos parameters $t_a$, the mirror Landau-Ginzburg model
is described by twisted chiral multiplets
$\sigma_a$, $Y_i$, and the superpotential
\begin{equation}
W \: = \: \sum_{a=1}^k \sigma_a \left( \sum_{i=1}^n \rho^a_i Y_i \: - \: t_a
\right) \: + \: \sum_{i=1}^n \exp\left( - Y_i \right).
\end{equation}
Operators in the mirror and the original gauge theory
 are related by the operator mirror
map 
\begin{equation}
\exp\left( - Y_i \right) \: \leftrightarrow \: \sum_{a} \rho_i^a \sigma_a,
\end{equation}
derived from the superpotential above (see e.g.
\cite{Gu:2017nye,Gu:2018fpm} for details).

This Landau-Ginzburg model is mirror in the sense that classical
computations in the B-twisted Landau-Ginzburg model reproduce
quantum computations in the A-twist of the gauge theory.
For one example, the axial $U(1)_A$ anomaly of the original gauge theory
appears as a classical obstruction to the existence of the
corresponding symmetry in the mirror theory, specifically
\begin{equation}
Y_i \: \mapsto \: Y_i - 2 i \alpha, 
\: \: \:
\sigma_a \: \mapsto \: \sigma_a \exp\left( + 2 i \alpha \right),
\end{equation}
where $\alpha$ parametrizes the symmetry,
at the same time that the superspace coordinates $\theta$ get phase
factors.  This symmetry has 
a classical obstruction
which can be resolved if one shifts the $\theta$ angle.
For another example, the quantum cohomology ring relations of the original
gauge theory are encoded in the classical critical locus of the 
mirror Landau-Ginzburg model.  For a third example,
integrating out the $Y$ fields returns the twisted one-loop effective
superpotential of the original A-twisted gauge theory.
More systematically, all correlation
functions of the original gauge theory, including quantum effects,
are reproduced from classical contributions to correlation functions
in the mirror B-twisted Landau-Ginzburg model.  In fact, more can be
said -- for example, open string sectors mirror in the expected
fashion -- but in this paper we focus on computations that have
heterotic analogues.

We have only described the mirror in the case that the original gauge
theory has no superpotential, but this description is straightforward
to modify in the presence of a superpotential.  Specifically, if the
original theory has a superpotential, then some of the chiral superfields
in the original gauge theory have nonzero R-charges.  In such a case,
we take the corresponding fundamental field in the mirror to be not $Y$
but instead $\exp( - (r/2) Y )$, where $r$ denotes the $r$ charge,
and the mirror also has a ${\mathbb Z}_{2/r}$ orbifold acting by phases
on that field.  (Twistability of the original theory restricts allowed
R-charges to $r \in \{0, 1, 2\}$, as discussed in \cite[section 2]{Gu:2018fpm}.)

The nonabelian extension proposed in \cite{Gu:2018fpm} followed
the same pattern, proposing a (B-twisted) Landau-Ginzburg mirror
to (nonabelian) A-twisted two-dimensional (2,2) supersymmetric
gauge theories, in which quantum effects in the A-twisted theory
are realized classically in the B-twisted mirror, which reduces to
\cite{Hori:2000kt} for abelian gauge theories.  Briefly,
for a $G$-gauge theory with $n$ chiral superfields in some
(typically reducible) representation of $G$, and Fayet-Iliopoulos
parameters $t_a$, $a \in \{1, \cdots, {\rm rank}\, G\}$,
the mirror Landau-Ginzburg model is defined by (a Weyl-group orbifold
of) twisted chiral
superfields
$\sigma_a$, $Y_i$ ($i \in \{1, \cdots, n \}$), and
$X_{\tilde{\mu}}$, the latter corresponding to nonzero roots of the
Lie algebra ${\mathfrak g}$ of the gauge group $G$, and a superpotential
\begin{equation}
W \: = \: \sum_{a=1}^{{\rm rank}\, G} \sigma_a \left(
\sum_{i=1}^n \rho_i^a Y_i \: - \: \sum_{\tilde{\mu}} \alpha^a_{\tilde{\mu}}
\ln X_{\tilde{\mu}} \: - \: t_a \right) \: + \:
\sum_{i=1}^n \exp\left(-Y_i \right) \: + \:
\sum_{\tilde{\mu}} X_{\tilde{\mu}}.
\end{equation}
Operators in the mirror and the original gauge theory
 are related by the operator mirror 
map 
\begin{equation}  \label{22-op-mirror}
\exp\left( - Y_i \right) \: \leftrightarrow \: \sum_{a} \rho_i^a \sigma_a, 
\: \: \:
X_{\tilde{\mu}} \: \leftrightarrow \: \sum_a \alpha^a_{\tilde{\mu}} \sigma_a,
\end{equation}
derived from the superpotential above (see e.g.
\cite{Gu:2017nye,Gu:2018fpm} for details).
The new ingredients, relative to the abelian case, are the fields
$X_{\tilde{\mu}}$, corresponding to nonzero roots of the Lie algebra
${\mathfrak g}$ of $G$, and the Weyl orbifold.

This proposal necessarily possesses all the same properties as 
the Hori-Vafa proposal, as well as some new ones.  For one example,
the axial anomaly of the original gauge theory is realized in the
mirror again as an obstruction to a classical symmetry, specifically
\begin{equation}  \label{eq:22-axial-mirror}
Y_i \: \mapsto \: Y_i - 2 i \alpha,
\: \: \:
X_{\tilde{\mu}} \: \mapsto \: X_{\tilde{\mu}} \exp\left( + 2 i \alpha \right),
\: \: \:
\sigma_a \: \mapsto \: \sigma_a \exp\left( + 2 i \alpha \right),
\end{equation}
where $\alpha$ parametrizes the symmetry, and the superspace
coordinates $\theta$ get phase factors.
The classical obstruction to this symmetry
can be
cured with a theta angle shift, just as in the abelian case.
For another example, quantum cohomology ring relations as well as
Coulomb branch relations (analogues of quantum cohomology relations in
cases where the Higgs branch has no weak-coupling limit, because of no
continuously-variable Fayet-Iliopoulos parameter) arising from
quantum corrections in the original gauge theory are realized
classically in the mirror as critical locus relations.
For a third example, integrating out the $X$ and $Y$ fields in the
mirror reproduces
the twisted one-loop effective superpotential of the original gauge theory.
More systematically, all correlation functions of the original gauge
theory, including quantum effects, are reproduced from classical contributions
to correlation functions in the mirror B-twisted Landau-Ginzburg model.

Nonabelian cases also possess a few additional properties.  For one example,
Coulomb branch vacua in a nonabelian two-dimensional gauge theory
are partly defined by `excluded loci,' constraining the $\sigma$ fields.
For example, in a $U(k)$ gauge theory with fundamental-valued matter,
for $a \neq b$, $\sigma_a \neq \sigma_b$.  (One way to understand this is
from supersymmetric localization, where the integration measure has a factor
proportional to $(\sigma_a - \sigma_b)^2$, which removes contributions from
coincident pairs of $\sigma$'s.)  One of the challenges in finding
a nonabelian mirror, one of the constraints on a possible mirror,
is to reproduce that excluded locus in the classical
physics of the B-twisted Landau-Ginzburg model.  Now, realizing a 
closed condition, such as a restriction to a subvariety, is relatively
straightforward, following the pattern described in
\cite{Witten:1993yc}.  The excluded locus condition above,
however, is an open condition, defining an open subset of the Coulomb
branch.  In the mirror proposal in \cite{Gu:2018fpm}, the excluded
locus condition is mirror to poles in the mirror superpotential.
For example, in the case of a $U(k)$ gauge theory with fundamentals,
the mirror theory has a field $X_{\mu \nu}$ which is mirror to
the difference $\sigma_{\mu} - \sigma_{\nu}$, and the superpotential
has a pole where $X_{\mu \nu} = 0$, implying that $\sigma_{\mu}$
must be distinct from $\sigma_{\nu}$.  In more general cases,
the excluded loci can be considerably more intricate, and one of the
checks performed in \cite{Gu:2018fpm} 
was to verify that the classical physics of the mirror did correctly
reproduce those excluded loci.

Numerous other consistency tests have also been performed.
For example, in the case of the two-dimensional gauge theory describing
Grassmannians $G(k,n)$, integrating out the $X$ fields reproduces a
proposal of \cite{Hori:2000kt} for the mirror to a Grassmannian.
In \cite{Hori:2000kt}, the proposal had factors of the form
\begin{equation}
\prod_{a < b} \left( \sigma_a - \sigma_b \right)^2
\end{equation}
in the integration measure, whose possible origin in a local field
theory was rather unclear, but becomes much more clear in the
mirror of \cite{Gu:2018fpm}.

\section{Proposal for (0,2) supersymmetric gauge theories}
\label{sect:proposal}

In this section, we will describe our ansatz for mirrors to (0,2) 
supersymmetric\footnote{For 
introductions to (0,2) GLSMs and (0,2) Landau-Ginzburg models,
we recommend \cite{Distler:1993mk,Distler:1995mi}.
}
GLSMs
which are deformations of (2,2) supersymmetric GLSMs, relating
the (0,2) supersymmetric analogue of the A-twist of the original
gauge theory (known\footnote{
For an introduction to the A/2 and B/2 twists, we refer the reader
to e.g. \cite{Katz:2004nn,Melnikov:2012hk,Melnikov:2019tpl}.
} as the A/2-twist) to the (0,2) supersymmetric
analogue of the B-twist (known as the B/2-twist) of the
mirror Landau-Ginzburg model.
Our ansatz will apply to both abelian and nonabelian theories obtained
as deformations of (2,2) supersymmetric theories, but with
a restriction on the allowed deformations, a restriction
on the form of the functions $E = \overline{D}_+ \Psi$, which
we shall describe in a moment.  

Our ansatz will follow the same pattern and have the same basic properties
as the other gauge theory mirrors discussed in the previous section.
For example, it will have the same symmetries, realizing classically
any anomalies of the original theory, as we shall see later in this
section.
For another example, quantum sheaf cohomology ring relations arising from
quantum corrections in the original (0,2) supersymmetric gauge theory are
realized classically in the mirror as critical locus relations, just
as in the (2,2) supersymmetric models, as we shall see explicitly later
in this section.  
For a third example, integrating
out the $X$ and $Y$ fields in the mirror reproduces the twisted one-loop
effective superpotential of the original gauge theory, just as in
(2,2) supersymmetric theories, as we discuss
in section~\ref{sect:just:int-out}.  More systematically, all (topological)
correlation functions of the original gauge theory, including quantum
effects, are reproduced from classical contributions to correlation functions
in the mirror B/2-twisted Landau-Ginzburg model, just as in 
(2,2) supersymmetric theories, as we discuss
in section~\ref{sect:just:corr}.

For simplicity, in this section we will
assume the original gauge theory has no superpotential, and will discuss
mirrors to theories with (0,2) superpotentials in 
section~\ref{sect:hypersurfaces}.  We do not claim a physical proof of this
proposal, though in later sections we will provide numerous consistency tests.

We will consider (0,2) theories that are deformations of (2,2) theories.
Now, (2,2) supersymmetric multiplets are equivalent to pairs of
(0,2) supersymmetric multiplets.  For example, a (2,2) supersymmetric
chiral superfield $\Phi$ is equivalent to a pair of (0,2)
supersymmetric multiplets:
\begin{itemize}
\item a (0,2) supersymmetric chiral multiplet $\Phi$,
\item a (0,2) supersymmetric Fermi multiplet $\Psi$,
with $\overline{D}_+ \Psi$ a holomorphic function of chiral superfields.
\end{itemize}
On the (2,2) locus, $\overline{D}_+ \Psi$ is uniquely specified.

We will consider
(0,2) deformations encoded in $\overline{D}_+ \Psi$, deforming
this function to a more general holomorphic function of the 
chiral superfields (and breaking (2,2) supersymmetry to (0,2)).
Specifically, we consider deformations obeying the following two
constraints:
\begin{itemize}
\item We assume that $\overline{D}_+ \Psi$ is linear in chiral superfields,
rather than a more general holomorphic function of chiral superfields.
This may
sound very restrictive, but in fact, it has been argued
that only linear terms contribute to A/2-twisted GLSMs\footnote{
For A/2-twisted nonlinear sigma models, this story is not settled,
not least because we know of no simple way to distinguish the UV linear
from UV nonlinear deformations in the IR.
} -- nonlinear terms are
irrelevant.  (This was conjectured in \cite{McOrist:2008ji}[section 3.5],
\cite{Kreuzer:2010ph}[section A.3], and rigorously 
proven in \cite{Donagi:2011uz,Donagi:2011va} for
abelian GLSMs. It also is a consequence of supersymmetric localization
\cite{Closset:2015ohf}, and see in addition \cite{Donagi:2014koa}[appendix A].)
\item We assume that $\overline{D}_+ \Psi$ is also diagonal,
meaning, for theories which are deformations of (2,2) theories, that
for any Fermi superfield $\Psi$, $\overline{D}_+ \Psi$ is proportional to the
chiral superfield with which it is partnered on the (2,2) locus.
\end{itemize}
On the (2,2) locus, the $\overline{D}_+ \Psi$ are both linear and diagonal,
and there exist nontrivial (0,2) deformations which are also linear and
diagonal.
The constraints above, that $\overline{D}_+ \Psi$ be both linear and
diagonal, imply the form
\begin{equation}
\overline{D}_+ \Psi_{i} =  E_{i}(\sigma) \Phi_{i}.
\end{equation}
This form is not the most general possible (0,2) deformation, 
but nevertheless still allows for nontrivial deformations,
and in any event is the most general possible deformation for which
we have been able to find a mirror construction that obeys all
consistency constraints.

Now that we have stated the restrictions, we give the proposal.
Let us consider
a (0,2) GLSM with connected\footnote{
It is very straightforward to extend this proposal
to $O(k)$ gauge theories in the
same fashion as the (2,2) case, discussed in \cite{Gu:2019zkw}, 
but we shall not
discuss any examples of $O(k)$ (0,2) mirrors in this paper.
} gauge group $G$ of dimension $n$ and rank $r$, chiral fields $\Phi_i$ and 
Fermi fields $\Psi_i$ in a (possibly reducible)
representation $R$ for $i=1, \cdots, N= {\rm dim}\, R$. 
If $\mathcal{W}$ is the Weyl group of $G$, then the proposed mirror theory is a $\mathcal{W}$-orbifold of a (0,2) Landau-Ginzburg model given by the following matter fields:
\begin{itemize}
\item $r$ chiral fields $\sigma_a$ and $r$ Fermi fields $\Upsilon_a$, $a=1,\cdots,r$,
\item chiral fields $Y_{i}$ and Fermi fields $F_{i}$ where $i=1,\cdots,N$,
\item $n-r$ chiral fields $X_{\tilde{\mu}}$ and $n-r$ Fermi fields 
$\Lambda_{\tilde{\mu}}$,
\end{itemize}
following the same pattern as the (2,2) nonabelian mirror
proposal \cite{Gu:2018fpm}.

For linear and diagonal $\overline{D}_+ \Psi$ as above,
the proposed (0,2) superpotential of the mirror Landau-Ginzburg orbifold is
\begin{equation}\label{superpotential}
\begin{split}
W=&\sum_{a=1}^r \Upsilon_a \left( \sum_{i=1}^N  \rho^a_i Y_{i}
\: - \: \sum_{\tilde{\mu}=1}^{n-r} \alpha_{\tilde{\mu}}^a \ln X_{\tilde{\mu}} 
\: - \: t_a \right)
\\
&  + \sum_{i=1}^N  F_{i}\left( E_{i}(\sigma)-\exp(-Y_{i})\right)
\: + \:
\sum_{\tilde{\mu}=1}^{n-r}\Lambda_{\tilde{\mu}}\left(1 \: - \:
\sum_{a=1}^r \sigma_a \alpha_{\tilde{\mu}}^a X_{\tilde{\mu}}^{-1} \right),
\end{split}
\end{equation}
where $\rho_i^a$ is the $a$-th component of the weight $\rho_i$ of
representation $R$,
and $\alpha_{\tilde{\mu}}$, $\tilde{\mu}=1, \cdots, n-r$ are the roots of $G$. 

In later sections, we will slightly modify the index structure above,
to be more convenient in each case, just as in
\cite{Gu:2018fpm,Chen:2018wep,Gu:2019zkw}.  For example, if the
matter representation $R$ consists
of multiple fundamentals, we will break $i$ into separate color and flavor
indices.

The Weyl orbifold group acts on the superpotential above in essentially
the same form as discussed in detail in
\cite{Gu:2018fpm,Chen:2018wep,Gu:2019zkw}, so we will be brief.
In broad brushstrokes, the orbifold group acts by a combination of
exchanging fields and multiplying by signs.  In the present case,
such actions happen on pairs $(Y_i, F_i)$, 
$(X_{\tilde{\mu}},\Lambda_{\tilde{\mu}})$, $(\sigma_a, \Upsilon_a)$
simultaneously.  For example, if $Y_i$ is swapped with $Y_j$,
then simultaneously $F_i$ is swapped with $F_j$.  If $Y_i$ is multiplied
by a sign, then simultaneously $F_i$ is multiplied by a sign.
It is then straightforward to show that the superpotential above is
invariant under the orbifold group, following the same arguments as
in \cite{Gu:2018fpm,Chen:2018wep,Gu:2019zkw}.

Furthermore, because the $\Lambda_{\tilde{\mu}}$ terms have the same form
as on the (2,2) locus, the part of the
excluded locus corresponding to $X_{\tilde{\mu}}$ poles
is the same as on the (2,2) locus,
and so, for mirrors to connected gauge groups, the fixed points of the Weyl
orbifold do not intersect non-excluded critical loci.
In passing, another part of the excluded locus is defined by the
fact that $\exp(-Y)$ is nonzero for finite $Y$, and that part of the excluded
locus will change as the $\exp(-Y)$'s are now determined by the $E$'s.

Most of the superpotential above is simply the (0,2) version of the
(2,2) mirrors of \cite{Hori:2000kt,Gu:2018fpm,Chen:2018wep,Gu:2019zkw}, 
with the exception of the $F E$ terms
in the second line.  For a (2,2) supersymmetric mirror, each of those
$E$'s would be 
\begin{equation}
E_{i}(\sigma) \: = \: \sum_{a=1}^r \rho_i^a \sigma_a.
\end{equation}
Allowing for more general $E$'s encodes the (0,2) deformation.  We should
also observe that in the original (0,2) gauge theory, those $E$'s are not
in the superpotential; the fact that they appear in the mirror (0,2)
superpotential is as one expects for mirror symmetry.

Just as in \cite{Gu:2018fpm}, we omit the K\"ahler potential from our
ansatz, partly because it is not pertinent to the tests we will perform.
For abelian (0,2) GLSMs, detailed discussions of dualities and corresponding
K\"ahler potentials can be found in \cite{Adams:2003zy}.

The constraints implied by the Fermi fields imply the operator
mirror map
\begin{eqnarray}\label{mirror map}
\exp(-Y_{i}) & = & E_{i}(\sigma),
\\
X_{\tilde{\mu}} & = & \sum_{a=1}^r \alpha^a_{\tilde{\mu}} \sigma_a,
\end{eqnarray}
a precise analogue of the operator mirror map~(\ref{22-op-mirror})
in the (2,2) case,
as well as the constraints
\begin{equation}
 \sum_{i=1}^N  \rho^a_i Y_{i}
\: - \: \sum_{\tilde{\mu}=1}^{n-r} \alpha_{\tilde{\mu}}^a \ln X_{\tilde{\mu}} 
\: = \: t_a.
\end{equation}
Exponentiating the constraints and applying the operator mirror map,
we get the relations
\begin{equation}
\left[ \prod_i E_i(\sigma)^{ \rho_i^a } \right]
\left[ \prod_{\tilde{\mu}} X_{\tilde{\mu}}^{ \alpha^a_{\tilde{\mu}} }
\right] \: = \: q_a.
\end{equation}
Just as in the (2,2) case \cite[section 3.3]{Gu:2018fpm}, 
and as we will see in more detail in
section~\ref{sec:justification}, the factor
\begin{equation}
\left[ \prod_{\tilde{\mu}} X_{\tilde{\mu}}^{ \alpha^a_{\tilde{\mu}} }
\right]
\end{equation}
just contributes a phase, so that these relations reduce to
\begin{equation}\label{qsc}
\prod_i E_{i}(\sigma)^{\rho^a_i} = \tilde{q}_a,
\end{equation}
for suitably phase-shifted $\tilde{q}_a \propto q_a$,
which are precisely the quantum sheaf cohomology relations for
these theories (see {\it e.g.} \cite{Closset:2015ohf}).
Thus, as expected, the quantum sheaf cohomology ring relations of the
original theory are realized classically in the mirror, just as
in (2,2) supersymmetric mirrors.

The right-chiral $U(1)_R$ symmetry of the original gauge theory
is realized in the mirror as
\begin{equation}
\begin{array}{c}
Y_i \: \mapsto \: Y_i - i \alpha,
\: \: \:
F_i \: \mapsto \: F_i \: \mbox{ (invariant) },
\\
X_{\tilde{\mu}} \: \mapsto \: X_{\tilde{\mu}} \exp\left( + i \alpha \right),
\: \: \:
\Lambda_{\tilde{\mu}} \: \mapsto \: \Lambda_{\tilde{\mu}} \exp\left(
+ i \alpha \right),
\\
\sigma_a \: \mapsto \: \sigma_a \exp\left( + i \alpha \right),
\: \: \:
\Upsilon_a \: \mapsto \: \Upsilon_a \exp\left( + i \alpha \right),
\end{array}
\end{equation}
and with a corresponding phase rotation of the superspace coordinates,
where $\alpha$ parametrizes the symmetry, following exactly the same
pattern as equation~(\ref{eq:22-axial-mirror}) for the (2,2) mirror,
and with the same result:  the axial anomaly of the original
gauge theory is mirror to a classical obstruction that can be cured
by a shift of the (mirror to the) theta angle.

The left-chiral $U(1)$ symmetry (an R-symmetry on the (2,2) locus) of the
original gauge theory is realized in the mirror as
\begin{equation}
\begin{array}{c}
Y_i \: \mapsto \: Y_i - i \alpha,
\: \: \:
F_i \: \mapsto \: F_i\exp\left( - i \alpha \right),
\\
X_{\tilde{\mu}} \: \mapsto \: X_{\tilde{\mu}} \exp\left( + i \alpha \right),
\: \: \:
\Lambda_{\tilde{\mu}} \: \mapsto \: \Lambda_{\tilde{\mu}} \:
\mbox{ (invariant) },
\\
\sigma_a \: \mapsto \: \sigma_a \exp\left( + i \alpha \right),
\: \: \:
\Upsilon_a \: \mapsto \: \Upsilon_a \: \mbox{ (invariant) },
\end{array}
\end{equation}
where $\alpha$ parametrizes the symmetry, and with no phase rotation of
the superspace coordinates.  As in the right-chiral case,
the anomaly of the original theory is realized in the mirror by a
classical obstruction that can be cured by a shift of the (mirror to the)
theta angle.

\section{Justification}\label{sec:justification}

We saw in the previous section that the proposed (0,2) mirror possesses
many of the desired properties of a mirror:  it realizes classically
the quantum sheaf cohomology ring relations of the original theory,
and it has the same symmetries, realizing anomalies classically in the 
mirror.

In this section, we will provide further general tests of the (0,2)
mirror proposal of the previous section.  Specifically, 
we will reproduce the one-loop effective (0,2)
superpotential of \cite{McOrist:2008ji} and also argue how correlation
functions in these theories reproduce those of the original gauge
theories, in cases in which vacua are isolated.
Our arguments in this section will be somewhat formal, but in concrete examples
in later sections we will verify these properties explicitly.

\subsection{Integrate out fields}
\label{sect:just:int-out}

In this section, we will integrate out fields and recover the
one-loop effective superpotential of the original gauge
theory, a standard property of (2,2) gauge theory mirrors that also
holds in this (0,2) supersymmetric mirror proposal.

First, following \cite{Gu:2018fpm}, to better understand the properties
of this theory, we integrate out the fields $X_{\tilde{\mu}}$ and
$\Lambda_{\tilde{\mu}}$.  This is an option because they have nonzero
masses; phrased simply,
\begin{equation}
\frac{ \partial^2 W}{\partial \Lambda_{\tilde{\mu}} \partial
X_{\tilde{\nu}} } \: = \: \sum_a \frac{ \sigma_a \alpha^a_{\tilde{\mu}} }{
X_{\tilde{\mu}}^2 } \delta_{ \tilde{\mu} \tilde{\nu} },
\end{equation}
whose zero locus defines part of the excluded locus,
as explained in \cite{Gu:2018fpm}.
The Hessian of $X_{\tilde{\mu}}$ is
\begin{equation}
H_X \: = \: \prod_{\tilde{\mu}} \left( \sum_{a=1}^r \sigma_a \alpha_\mu^a \right)^{-1},
\end{equation}
which, when integrating out $X_{\tilde{\mu}}$,
$\Lambda_{\tilde{\mu}}$, 
generates a factor in the path integral measure which vanishes
along the excluded locus, exactly the same as in (2,2)
mirrors \cite{Gu:2018fpm}.
The equations of motion of $X_{\tilde{\mu}}$ are
\begin{equation}
X_{\tilde{\mu}} \: = \: \sum_{a=1}^r \sigma_a \alpha^a_{\tilde{\mu}}.
\end{equation}
Therefore, integrating out $X_{\tilde{\mu}}$ and $\Lambda_{\tilde{\mu}}$ 
amounts to eliminating the terms proportional to 
$\Lambda_{\tilde{\mu}}$ and $\ln X_{\tilde{\mu}}$ in \eqref{superpotential} 
and shifting the FI parameters $t_a$ to $\tilde{t}_a$, just as happens
in (2,2) mirrors \cite{Gu:2018fpm}, reproducing a phase discussed
in \cite[section 10]{Hori:2013ika}. 
For example, for each $U(k)$ factor of the gauge group,
\begin{equation}
\alpha_{\tilde{\mu}}^a \: = \: \alpha_{bc}^a \: = \: \delta_c^a-\delta_b^a
\end{equation}
for $a,b,c=1,\cdots,k$ and $b \neq c$ and thus
\begin{equation}
\sum_{\tilde{\mu}} \alpha_{\tilde{\mu}}^a \ln X_{\tilde{\mu}} 
\: =  \:
\sum_{b \neq c} \alpha_{bc}^a \ln\left( \sigma_c - \sigma_b \right) 
\: = \: 
\sum_{b \neq a} \ln \left( \frac{\sigma_a-\sigma_b}{\sigma_b-\sigma_a} \right) 
\: = \: (k-1) \pi i
\end{equation}
from the equation of motion.

Therefore, after integrating out $X_{\tilde{\mu}}$ and $\Lambda_{\tilde{\mu}}$,
the superpotential \eqref{superpotential} reduces to
\begin{equation}\label{superpotential-}
\tilde{W}
\: = \:
\sum_{a=1}^r \Upsilon_a \left( \sum_{i=1}^N  \rho^a_i Y_{i} - \tilde{t}_a \right)
\: + \: \sum_{i=1}^N F_{i}\left( E_{i}(\sigma)-\exp(-Y_{i})\right)
\end{equation}
through a redefinition $\tilde{t}_a$ of $t_a$.
The equations of motion of $\sigma_a$ and $Y_{i}$ derived from \eqref{superpotential-} then gives the mirror map \eqref{mirror map} and the expected quantum sheaf cohomology relations \eqref{qsc}.

Let us now also integrate out $(Y_i, F_i)$.  We will recover the one-loop
effective superpotential of the original gauge theory on the Coulomb
branch, just as happens in (2,2) supersymmetric mirrors. 
As before, it is legitimate to do so because these fields have nonzero mass:
\begin{equation}
\frac{\partial^2 W}{\partial Y_i \partial F_j} \: = \:
\delta_{ij} \exp\left( - Y_i \right),
\end{equation}
whose zero locus defines part of the excluded locus,
as explained in \cite{Gu:2018fpm}.
From the superpotential above, we find 
\begin{equation}
\exp\left( - Y_i \right) \: = \: E_i(\sigma),
\end{equation}
or simply
\begin{equation}
Y_i \: = \: - \ln E_i(\sigma),
\end{equation}
and plugging back in we recover
\begin{equation}
\tilde{W}
\: = \:
\sum_{a=1}^r \Upsilon_a \left(
- \sum_{i=1}^N \rho^a_i \ln E_i(\sigma) - \tilde{t}_a \right).
\end{equation}
This matches \cite[equ'ns (3.22)-(3.23)]{McOrist:2008ji}.
Thus, we see that integrating out fields recovers the one-loop
effective superpotential along the Coulomb branch,
exactly as happens in (2,2) supersymmetric gauge theory mirrors.

\subsection{Correlation functions}
\label{sect:just:corr}

In this section, we will compare correlation functions in the B/2-twisted
Landau-Ginzburg model just defined~\eqref{superpotential} with corresponding
A/2 model correlation functions, in cases with isolated Coulomb branch
vacua, and along the way, recover the
one-loop effective (0,2) superpotential of 
\cite{McOrist:2008ji} along the Coulomb branch.  

Now, for a (0,2) superpotential of
the form $W = F^i J_i$ with isolated vacua, correlation functions are
schematically of the form \cite{Melnikov:2007xi}
\begin{equation}
\langle f \rangle \: = \: \sum_{\rm vacua} \frac{ f }{ \det \partial_i J_j },
\end{equation}
closely related to formulas for correlation functions in (2,2) Landau-Ginzburg
models involving determinants of matrices of second derivatives of the
superpotential.
Thus, we need to compute some analogues of Hessians.

The Hessian of $Y_{i}$ is
\[
H_Y = \prod_i  \exp(-Y_{i}) = \prod_i E_{i}(\sigma),
\]
which is nonzero at generic points on the Coulomb branch. 
From~\eqref{mirror map}, integrating out $Y_{i}$ and $F_{i}$ reduces~\eqref{superpotential-} to
\begin{equation}\label{superpotential--}
W_{\rm eff} \: = \: \sum_{a=1}^r \Upsilon_a J^a_{\rm eff}
\: = \: 
\sum_{a=1}^r \Upsilon_a \left( -\sum_{i=1}^N  \rho^a_i \ln E_{i}(\sigma) - \tilde{t}_a \right),
\end{equation}
which is the same as the effective superpotential on the Coulomb branch of the original GLSM. 
Consequently, assuming isolated vacua,
for any operator $\mathcal{O}(\sigma)$, the B/2 correlation functions of our proposed Landau-Ginzburg mirror are \cite{Melnikov:2007xi}
\begin{eqnarray}
\langle\mathcal{O}(\sigma)\rangle & = & \frac{1}{|\mathcal{W}|} \sum_{J^a_{\rm eff}=0} \frac{\mathcal{O}(\sigma)}{\left( \det_{a,b} \partial_b J^a_{\rm eff} \right) H_X H_Y},
\\
& = & \frac{1}{|\mathcal{W}|} \sum_{J^a_{\rm eff}=0} \frac{\mathcal{O}(\sigma) \prod_{\tilde{\mu}} \left( \sum_{a=1}^r \sigma_a \alpha_{\tilde{\mu}}^a \right)}{\left( \det_{a,b} \partial_b J^a_{\rm eff} \right) \left(\prod_i E_{i}(\sigma)\right)},
\end{eqnarray}
which is the same as the A/2 correlation function computed from the original 
GLSM \cite[equ'n (3.63)]{Closset:2015ohf}.
(The factor of $1/| {\mathcal W}|$ reflects the Weyl orbifold, which
acts freely on the critical locus, as in \cite{Gu:2018fpm}, so that twisted
sectors do not enter this computation, at least for mirrors to theories
with connected gauge groups.)

\section{Specialization to abelian theories}
\label{sect:specialize}

Let's consider a GLSM with gauge group $U(1)^r$. 
The chiral field $\Phi_i$ and Fermi field $\Psi_i$ have charge $Q^a_i$ under
the $a$-th $U(1)$, for $i=1, \cdots, N$. Assuming linear and diagonal
(0,2) deformations, as discussed before, these fields satisfy
\begin{equation}
\overline{D}_+ \Psi_i \: = \: \sum_{a=1}^r E_i^a \sigma_a \Phi_i,
\end{equation}
where $E_i^a=Q_i^a$ on the (2,2) locus.

In the abelian case, the fields $X_\mu$ and $\Lambda_\mu$ are absent in the mirror theory. 
The matter content of the mirror Landau-Ginzburg model thus consists of chiral fields $\sigma_a, Y_i$ and Fermi fields $\Upsilon_a, F_i$, $a=1,\cdots,r, i=1,\cdots,N$. The superpotential is
\begin{equation}
W \: = \:
\sum_{a=1}^r \Upsilon_a \left( \sum_{i=1}^N Q_i^a Y_i - t^a \right)
\: + \:
 \sum_{i=1}^N F_i \left( \sum_{a=1}^r E_i^a \sigma_a - \exp(-Y_i) \right).
\end{equation}
Specializing equation~(\ref{mirror map}), the operator mirror map in this
case is
\begin{equation}
\exp(-Y_i) \: = \: \sum_{a=1}^r E_i^a \sigma_a
\end{equation}
and the effective superpotential is
\begin{equation}
W_{\rm eff} \: = \: \sum_{a=1}^r \Upsilon_a J^a_{\rm eff}
\: = \: 
\sum_{a=1}^r \Upsilon_a \left( -\sum_{i=1}^N Q_i^a \ln \left(\sum_{b=1}^r E_i^b \sigma_b\right) - t^a \right),
\end{equation}
which reproduces the expected correlation functions
\begin{equation}
\langle \mathcal{O}(\sigma) \rangle \: = \: \sum_{J^a_{\rm eff}=0} \frac{\mathcal{O}(\sigma)}{\left( \det_{a,b} \partial_b J^a_{\rm eff} \right) H_Y},
\end{equation}
where
\begin{equation}
H_Y \: = \: \prod_{i=1}^N \left(\sum_{a=1}^r E_i^a \sigma_a \right).
\end{equation}

\section{Comparison to previous abelian proposal}
\label{sect:previous}

A proposal was made for a systematic mirror construction in abelian
(0,2) GLSMs in \cite{Gu:2017nye}.  The proposal of this paper both
generalizes and simplifies the proposal given there.  In this section,
we will explicitly relate our ansatz to that discussed there.
(Special cases have already been discussed, in
sections~\ref{sect:pnpm:compare} and \ref{sect:hirzebruch}.) 

Briefly, the proposal in \cite{Gu:2017nye} considered abelian (0,2) GLSMs
with $E$'s that are both linear and diagonal, as here, but with two
additional restrictions:
\begin{itemize}
\item To compute the mirror, one picked an invertible submatrix $S$ of the
charge matrix,
\item and the (0,2) deformations vanished for $E$'s corresponding to rows of
$S$.
\end{itemize}
The physics of the resulting mirror was independent of choices, but nevertheless
this was a more restrictive mirror than that given in this paper.

We will outline a derivation of the construction in
\cite{Gu:2017nye} from the mirror in this paper, but first, with
the benefit of hindsight, let us
outline in general terms how they are related.
\begin{itemize}
\item In the proposal of this paper, to generate a lower-energy Landau-Ginzburg
model, we may for example integrate out a subset of the $F$ Fermi fields,
and solve for the $\sigma_a$.  This procedure only works if the corresponding
submatrix of the $E$'s is invertible, and so, broadly speaking, corresponds
to a choice of invertible submatrix.
\item Assuming that the $E$ submatrix chosen above is the same as on the (2,2)
locus removes the necessity of keeping track of overall numerical factors
multiplying partition functions and correlation functions, the subtlety
discussed in {\it e.g.} subsection~\ref{sect:pnpm:first}.
\end{itemize}

Next, we shall outline a derivation of
the ansatz of \cite{Gu:2017nye} from the proposal of this paper.
First, they wrote their linear diagonal $\overline{D}_{+} \Psi_i$
in terms of deformations $B_{ij}$ off the (2,2) locus, as
\begin{equation}
E_i \: = \: \sum_j \sum_a \left( \delta_{ij} + B_{ij} \right) Q_i^a
\sigma_a.
\end{equation}
For these $E_i$, our ansatz~(\ref{superpotential}) can be written as
\begin{eqnarray}
W & = & \sum_{a=1}^r \Upsilon_a \left( \sum_{i=1}^N Q_i^a \sigma_a \: - \:
t_a \right)
\nonumber \\
& & \hspace*{0.5in} 
 \: + \:
\sum_{i=1}^N F_i \left( \sum_a Q_i^a \sigma_a \: + \:
 \sum_j B_{ij} Q^a_j \sigma_a
\: - \: \exp\left( - Y_i \right) \right).
\end{eqnarray}
Now, in the ansatz of \cite{Gu:2017nye}, one picks an invertible submatrix
$S$ of the charge matrix, and for $i$ corresponding to a column of $S$,
$B_{ij} = 0$.  As a result, for those $i$, the $F_i$ terms are simply
\begin{equation}
F_i \left( \sum_a Q^a_i \sigma_a \: - \: \exp\left( - Y_i \right) \right),
\end{equation}
and so we have a constraint that relates, for those $i$,
\begin{equation}
\sum_a Q^a_i \sigma_a \: = \: \exp\left( - Y_i \right),
\end{equation}
or equivalently, in the notation of \cite{Gu:2017nye},
\begin{equation}
\sum_a S^a_{i_S} \sigma_a \: = \: \exp\left( - Y_i \right).
\end{equation}
Solving for $\sigma_a$, we have
\begin{equation}
\sigma_a \: = \: \sum_{i_S} \left( S^{-1} \right)_{a i_S} 
\exp\left( - Y_{i_S} \right),
\end{equation}
and plugging back in, our (0,2) superpotential becomes
\begin{eqnarray}
W & = & \sum_{a=1}^r \Upsilon_a \left( \sum_{i=1}^N Q_i^a \sigma_a \: - \:
t_a \right)
\nonumber \\
& &  
 \: + \:
\sum_{i=1}^N F_i \left( \sum_a Q_i^a \sigma_a \: + \:
 \sum_{a, j, i_S} B_{ij} Q^a_j
 \left( S^{-1} \right)_{a i_S} 
\exp\left( - Y_{i_S} \right) 
\: - \: \exp\left( - Y_i \right) \right),
\end{eqnarray}
which is precisely the (0,2) superpotential of
\cite{Gu:2017nye}.

\section{Example:  ${\mathbb P}^n \times {\mathbb P}^m$}
\label{sect:pnpm}

So far we have given general arguments that expected properties of the
mirror always hold for this proposal:  anomalies of the original
theory are realized classically
in the mirror, quantum sheaf cohomology ring relations arise from
classical critical locus constraints, correlation functions match,
and integrating out fields returns the one-loop twisted effective
action of the original theory, all as expected for a gauge theory
mirror ala \cite{Hori:2000kt,Gu:2018fpm}.

Now, general arguments are well and good, but to make the discussion
more concrete, working through examples can also be helpful.
To that end, in this section we work through the first of several
examples, to see concretely how the mirror works in special cases.

\subsection{Setup}
\label{sect:pnpm:UV}

In this section we will compare to proposals for (0,2) mirrors to
${\mathbb P}^n \times {\mathbb P}^m$ with a deformation of the
tangent bundle, as discussed in \cite{Chen:2016tdd}.

In this case, a general deformation of the tangent bundle is
described as the cokernel
\begin{equation}
0 \: \longrightarrow \: {\cal O}^2 \: \stackrel{E}{\longrightarrow} \:
{\cal O}(1,0)^{n+1} \oplus {\cal O}(0,1)^{m+1} \:
\longrightarrow \: {\cal E} \: \longrightarrow \: 0,
\end{equation}
where 
\begin{equation}
E \: = \: \left[ \begin{array}{cc} Ax & B x \\ C y & D y \end{array} \right],
\end{equation}
where $x$, $y$ are vectors of homogeneous coordinates on
${\mathbb P}^n$, ${\mathbb P}^m$, respectively, and where $A$, $B$
are constant $(n+1) \times (n+1)$ matrices, and $C$, $D$ are
constant $(m+1) \times (m+1)$ matrices.  In this language,
the (2,2) locus corresponds for example to the case that $A$ and $D$
are identity matrices, and $B = 0 = C$.

Physically, in the corresponding (0,2) GLSM, we can write
\begin{equation}
\overline{D}_+ \Lambda_i \: = \: \left( A_{ij} \sigma +
B_{ij} \tilde{\sigma} \right) x_j,
\: \: \:
\overline{D}_+ \tilde{\Lambda}_j \: = \: \left( C_{jk} \sigma +
D_{jk} \tilde{\sigma} \right) y_k,
\end{equation}
and so we have 
\begin{equation}
E_{ij}(\sigma, \tilde{\sigma}) \: = \: (A \sigma + B \tilde{\sigma})_{ij},
\: \: \:
\tilde{E}_{jk}(\sigma, \tilde{\sigma}) \: = \:
(C \sigma + D \tilde{\sigma})_{jk}.
\end{equation}
The (0,2) mirror ansatz of this paper is only defined for
diagonal $E$'s,
so we shall assume the matrices $A$, $B$, $C$, $D$
are diagonal:
\begin{eqnarray}
A & = & {\rm diag}\left( a_0, \cdots, a_n \right),
\\
B & = & {\rm diag}\left( b_0, \cdots, b_n \right),
\\
C & = & {\rm diag}\left( c_0, \cdots, c_m \right),
\\
D & = & {\rm diag}\left( d_0, \cdots, d_m \right).
\end{eqnarray}
We also define
\begin{equation}
E_i(\sigma, \tilde{\sigma}) \: = \: a_i \sigma + b_i \tilde{\sigma},
\: \: \:
\tilde{E}_i(\sigma, \tilde{\sigma}) \: = \: c_i \sigma +
d_i \tilde{\sigma}.
\end{equation}

Following the (0,2) mirror ansatz given earlier, we take the (0,2) mirror
to be defined by the superpotential
\begin{eqnarray}
W & = & \Upsilon_1 \left( \sum_{i=1}^n Y_i \: - \: t_1 \right)
\: + \:
\Upsilon_2\left( \sum_{j=0}^m \tilde{Y}_j \: - \: t_2 \right)
\nonumber \\
& & 
\: + \: \sum_{i=0}^n F_i \left( E_i(\sigma, \tilde{\sigma}) \: - \:
\exp\left( - Y_i \right) \right)
\: + \:
\sum_{j=0}^m \tilde{F}_j \left( \tilde{E}_j(\sigma, \tilde{\sigma}) \: - \:
\exp\left( - \tilde{Y}_j \right) \right).
\label{eq:pnpm:mirror1}
\end{eqnarray}

As a first consistency test, let us verify that this produces the
quantum sheaf cohomology ring of ${\mathbb P}^n \times {\mathbb P}^m$.
First, we
integrate out the $\Upsilon_i$, which gives the usual constraints
\begin{equation}  \label{eq:pnpm:constr1}
\prod_{i=0}^n \exp\left( - Y_i \right) \: = \: q_1,
\: \: \:
\prod_{j=0}^m \exp\left( - \tilde{Y}_j \right) \: = \: q_2.
\end{equation}

Integrating out the $F_i$, $\tilde{F}_j$ gives the operator mirror maps
\begin{equation}
\exp\left( - Y_i \right) \: = \: E_i(\sigma, \tilde{\sigma}),
\: \: \:
\exp\left( - \tilde{Y}_j \right) \: = \: \tilde{E}_j(\sigma, \tilde{\sigma}),
\end{equation}
and combining these with the constraints~(\ref{eq:pnpm:constr1}),
one immediately has
\begin{equation}
\det(A \sigma + B \tilde{\sigma}) \: = \:
\prod_i E_i(\sigma, \tilde{\sigma}) \: = \: q_1,
\: \: \:
\det(C \sigma + D \tilde{\sigma}) \: = \:
\prod_j \tilde{E}_j(\sigma, \tilde{\sigma}) \: = \: q_2,
\end{equation}
which are precisely the quantum sheaf cohomology ring relations for
this model
\cite{McOrist:2007kp,McOrist:2008ji,Donagi:2011uz,Donagi:2011va,Closset:2015ohf}.

\subsection{Correlation functions in the UV}
\label{sect:p1p1:2pt:uv}

Before going on to integrate out some of the fields, 
let us take a moment to explicitly compute 
two-point B/2-model correlation functions in the case of the mirror to
${\mathbb P}^1 \times {\mathbb P}^1$.  (As we already know the chiral
ring matches that of the A/2 model, from the results of the immediately
preceding subsection, computing the two-point correlation
functions suffices to determine all of the B/2-model correlation functions.)

Correlation functions for the ${\mathbb P}^1 \times {\mathbb P}^1$
model were computed in \cite{Closset:2015ohf}[section 4.2].  We repeat
the highlights here for completeness.  The two-point correlation functions
have the form
\begin{equation}
\langle \sigma \sigma \rangle \: = \: - \frac{\Gamma_1}{\alpha},
\: \: \:
\langle \sigma \tilde{\sigma} \rangle \: = \: + \frac{\Delta}{\alpha},
\: \: \:
\langle \tilde{\sigma} \tilde{\sigma} \rangle \: = \: - \frac{\Gamma_2}{
\alpha},
\end{equation}
where
\begin{equation}
\begin{array}{lr}
\gamma_{AB} \: = \: \det(A+B) - \det A - \det B,
&
\gamma_{CD} \: = \: \det(C+D) - \det C - \det D,
\\
\Gamma_1 \: = \: \gamma_{AB} \det D - \gamma_{CD} \det B,
&
\Gamma_2 \: = \: \gamma_{CD} \det A - \gamma_{AB} \det C,
\\
\Delta \: = \: (\det A)(\det D) - (\det B)(\det C),
&
\alpha \: = \: \Delta^2 - \Gamma_1 \Gamma_2.
\end{array}
\end{equation}

We can compute correlation functions in the present mirror B/2-twisted
Landau-Ginzburg model with superpotential~(\ref{eq:pnpm:mirror1} 
using the methods of \cite{Melnikov:2007xi}.
Specializing to $n=m=1$, we have six functions $J_i$, corresponding
to the coefficients of $\Upsilon_{1,2}$, $F_{1,2}$, $\tilde{F}_{1,2}$,
and six fields $\sigma$, $\tilde{\sigma}$, $Y_{0,1}$,
$\tilde{Y}_{0,1}$.  The resulting matrix of derivatives $(\partial_i J_j)$
has the form
\begin{equation}
(\partial_i J_j) \: = \: \left[ \begin{array}{cccccc}
0 & 0 & 1 & 1 & 0 & 0 \\
0 & 0 & 0 & 0 & 1 & 1 \\
a_0 & b_0 & \exp\left( - Y_0 \right) & 0 & 0 & 0 \\
a_1 & b_1 & 0 & \exp\left( - Y_1 \right) & 0 & 0 \\
c_0 & d_0 & 0 & 0 & \exp\left( - \tilde{Y}_0 \right) & 0 \\
c_1 & d_1 & 0 & 0 & 0 & \exp\left( - \tilde{Y}_1 \right) 
\end{array} \right],
\end{equation}
and then correlation functions have the form
\begin{equation}
\langle f(\sigma,\tilde{\sigma}) \rangle \: = \: 
\sum_{ J=0 } \frac{ f(\sigma, \tilde{\sigma}) }{ \det (\partial_i J_j ) },
\end{equation}
where the sum is over the solutions of $\{ J_i = 0 \}$.
It is straightforward to compute that the resulting correlation functions
precisely match those listed above from the A/2 model 
\cite{Closset:2015ohf}[section 4.2].

\subsection{More nearly standard expressions}
\label{sect:pnpm:IR}

More nearly standard expressions for Landau-Ginzburg mirrors do not
involve $\sigma$ fields, so in this section, we shall integrate out these
fields to derive expressions for mirrors of a more nearly standard form.
We will encounter some interesting subtleties.

Specifically, some other expressions for possible (0,2) mirrors
to ${\mathbb P}^n \times {\mathbb P}^m$ are in 
\cite{Chen:2016tdd,Gu:2017nye}.  Those expressions have precisely $n$ $Y$'s
and $m$ $\tilde{Y}$'s, so we first integrate out the $\Upsilon_i$,
eliminating $Y_0$, $\tilde{Y}_0$:
\begin{equation}
\exp\left( - Y_0 \right) \: = \: q_1 \prod_{i=1}^n \exp\left( + Y_i \right),
\: \: \:
\exp\left( - \tilde{Y}_0 \right) \: = \: q_2 \prod_{j=1}^m
\exp\left( + \tilde{Y}_j \right).
\end{equation}

Next, we can either integrate out some of the Fermi fields $F_i$, 
$\tilde{F}_j$, and then integrate out $\sigma$'s, or we can integrate
out $\sigma$'s first, and then some of the Fermi fields.  
This order-of-operations
ambiguity
does not exist in (2,2) theories.  The results are independent of
choices, as one should expect,
but we illustrate both methods next, to illustrate various subtleties
in both the analysis and the normalization of the results.
In later analyses in this paper, we will be much more brief.

\subsubsection{First method}  \label{sect:pnpm:first}

Having integrating out the $\Upsilon_i$, our strategy in this approach
is to next integrate out some $F$, $\tilde{F}$ (as many as $\sigma$'s),
and then use the resulting constraints to eliminate $\sigma$'s.

The expressions in \cite{Chen:2016tdd,Gu:2017nye} have as many $F$'s as
$Y$'s, so we need to integrate out one $F$ and one $\tilde{F}$.  This will
mean solving for $\sigma$ and $\tilde{\sigma}$ in terms of other variables.
There are a number of ways to proceed, and indeed, one expects that
there will be many equivalent but different-looking expresions for
$\sigma$, $\tilde{\sigma}$ in terms of $Y_i$ and $\tilde{Y}_j$.
To pick one, we choose an index $i$ and $j$ such that the 
expressions we get from integrating out the corresponding $F$
and $\tilde{F}$, namely
\begin{equation}
\exp\left( - Y_i \right) \: = \: E_i(\sigma, \tilde{\sigma}),
\: \: \:
\exp\left( - \tilde{Y}_j \right) \: = \: \tilde{E}_j(\sigma, \tilde{\sigma}),
\end{equation}
can be inverted to solve for $\sigma$, $\tilde{\sigma}$ in terms of
$Y_i$, $\tilde{Y}_j$.  Put another way, using an index $I$ to denote
either $i$ or $j$, and writing, schematically,
\begin{equation}
E_I(\sigma,\tilde{\sigma}) \: = \: S_I^{\alpha} \sigma_{\alpha},
\end{equation}
we pick two indices $I$ such that the resulting $2 \times 2$ matrix
$S$ is invertible.  (Here we are deliberately making contact with the
notation used in \cite{Gu:2017nye}.)

Suppose, for example, that the two equations
\begin{equation}  \label{eq:pnpm:f0ft0:1}
\exp\left( - Y_0 \right) \: = \: E_0(\sigma, \tilde{\sigma}),
\: \: \:
\exp\left( - \tilde{Y}_0 \right) \: = \: \tilde{E}_0(\sigma, \tilde{\sigma}),
\end{equation}
can be inverted to solve for $\sigma$, $\tilde{\sigma}$.  Let us do this
explicitly, and examine the result.  From our earlier discussion,
\begin{equation}
E_0(\sigma, \tilde{\sigma}) \: = \: a_0 \sigma + b_0 \tilde{\sigma},
\: \: \:
\tilde{E}_0(\sigma, \tilde{\sigma}) \: = \: c_0 \sigma +
d_0 \tilde{\sigma}.
\end{equation}
Assuming that
\begin{equation}
\Delta_0 \: \equiv \: \det \left[ \begin{array}{cc}
a_0 & b_0 \\ c_0 & d_0 \end{array} \right] \: \neq \: 0,
\end{equation}
we first integrate out $F_0$, $\tilde{F}_0$ to get the 
constraints~(\ref{eq:pnpm:f0ft0:1}), and 
then these equations to find
\begin{eqnarray}
\sigma & = & \frac{1}{\Delta_0} \left( d_0 \exp\left( - Y_0 \right)
\: - \: b_0 \exp\left( - \tilde{Y}_0 \right) \right),
\\
\tilde{\sigma} & = & \frac{1}{\Delta_0} \left( - c_0 \exp\left( - Y_0 \right)
\: + \: a_0 \exp\left( - \tilde{Y}_0 \right) \right).
\end{eqnarray}

Then, after finally integrating out 
$\sigma$ and $\tilde{\sigma}$, the (0,2) superpotential
reduces to
\begin{eqnarray}
W & = & \sum_{i=1}^n F_i \left( a_i \sigma + b_i \tilde{\sigma} - 
\exp\left( - Y_i \right) \right)
\: + \: \sum_{j=1}^m \tilde{F}_j \left( c_j \sigma +
d_j \tilde{\sigma} - \exp\left( - \tilde{Y}_j \right) \right),
\\
& = &  \sum_{i=1}^n F_i \Biggl[ 
\frac{ \left( a_i d_0 - b_i c_0 \right) }{\Delta_0} 
q_1 \prod_{i'=1}^n \exp\left( + Y_{i'} \right)
\: + \:
\frac{ \left( - a_i b_0 + b_i a_0 \right)}{\Delta_0}
 q_2 \prod_{j'=1}^m \exp\left( +
\tilde{Y}_{j'} \right)
\nonumber \\
& & \hspace*{3.0in}
 \: - \: \exp\left( - Y_i \right) \Biggr]
\nonumber \\
& & 
\: + \:
 \sum_{j=1}^m \tilde{F}_j \Biggl[
\frac{ \left( c_j d_0 - d_j c_0 \right)}{\Delta_0}
 q_1 \prod_{i'=1}^n \exp\left( + Y_{i'} \right)
\: + \:
\frac{ \left( - c_j b_0 + d_j a_0 \right)}{\Delta_0}
 q_2 \prod_{j'=1}^m \exp\left(
+ \tilde{Y}_{j'} \right)
\nonumber \\
& & \hspace*{3.0in} \: - \: \exp\left( - \tilde{Y}_j \right)
\Biggr].    \label{eq:pnpm:first}
\end{eqnarray}

Before going on, there is a subtlety we should discuss, that will
become important when comparing correlation functions between the UV
and lower-energy theories.  Specifically, when we integrated out $\sigma$ and
$\tilde{\sigma}$, one effect is to multiply the path integral by
a constant.  Specifically, after integrating out 
$F_0$ and $\tilde{F}_0$, we had constraints which schematically appear
in the B/2 model path integral in the form
\begin{equation}
\int d \sigma d \tilde{\sigma} \,
\delta\left( a_0 \sigma + b_0 \tilde{\sigma} - \exp\left( - Y_0 \right) \right)
\delta\left( c_0 \sigma + d_0 \tilde{\sigma} - \exp\left( -
\tilde{Y}_0 \right) \right).
\end{equation}
Then, integrating over $\sigma$, $\tilde{\sigma}$ generates a factor
of 
\begin{equation}
\frac{1}{a_0 d_0 - b_0 c_0} \: = \: \frac{1}{\Delta_0}
\end{equation}
from the Jacobian.  This will multiply correlation functions in the lower-energy
theory,
and we will see later in subsection~\ref{sect:pnpm:corrfns:ir} 
that this will be required in order for
the lower-energy-theory's correlation functions to
match the UV correlation functions.

\subsubsection{Second method}

As a consistency test, and to illuminate the underlying methods,
we will now rederive the same result via a different approach.
Having integrated out the $\Upsilon_i$, our strategy in this approach
is to next integrate out the $\sigma_a$.  This will generate constraints
on the $F$, $\tilde{F}$, which we will use to write some in terms of the
others.  (This is the opposite order of operations from the previous approach.)

The result of this method will be an expression for the (0,2) mirror
that is not of the form described in \cite{Chen:2016tdd,Gu:2017nye}, 
and also does not respect
symmetries of the parametrization.  

We restrict to ${\mathbb P}^1 \times {\mathbb P}^1$ for simplicity.
Integrating out $\sigma_a$, we have the constraints
\begin{eqnarray}
\sum_{i=0}^n a_i F_i \: + \: \sum_{j=0}^m c_j \tilde{F}_j & = & 0,
\\
\sum_{i=0}^n b_i F_i \: + \: \sum_{j=0}^m d_j \tilde{F}_j & = & 0.
\end{eqnarray}
Solving for $F_0$, $\tilde{F}_0$, we find
\begin{eqnarray}
F_0 
& = &
- \frac{1}{\Delta_0} \left[ \sum_{i=1}^n (a_i d_0 - b_i c_0) F_i
\: + \:
\sum_{j=1}^m (c_j d_0 - c_0 d_j) \tilde{F}_j \right],
\\
\tilde{F}_0 
& = &
- \frac{1}{\Delta_0} \left[ \sum_{i=1}^n (a_0 b_i - b_0 a_i) F_i \: + \:
\sum_{j=1}^m (d_j a_0 - b_0 c_j) \tilde{F}_j \right]
\end{eqnarray}
where
\begin{equation}
\Delta_0 \: = \: a_0 d_0 - b_0 c_0.
\end{equation}

Plugging this back into the (0,2) superpotential, we have
\begin{eqnarray}
W & = &
- \sum_{i=0}^n F_i \exp\left( - Y_i \right) \: - \:
\sum_{j=0}^m \tilde{F}_j \exp\left( - \tilde{Y}_j \right),
\\
& = &
-  \sum_{i=1}^n F_i \Biggl[  \exp\left( - Y_i \right) \: - \:
\frac{(a_i d_0 - b_i c_0)}{\Delta_0} q_1 \prod_{k=1}^n \exp\left( + Y_k \right)
\nonumber \\
& & \hspace*{1.5in}
\: - \:
\frac{ (a_0 b_i - b_0 a_i)}{\Delta_0} q_2 \prod_{k=1}^m \exp\left( +
\tilde{Y}_k \right) \Biggr]
\nonumber \\
& &
- \sum_{j=1}^m \tilde{F}_j \Biggl[ \exp\left( - \tilde{Y}_j \right) \: - \:
\frac{ (c_j d_0 - c_0 d_j) }{\Delta_0} q_1 \prod_{k=1}^n \exp\left( + Y_k 
\right)
\nonumber \\
& & \hspace*{1.5in}
\: - \:
\frac{ (d_j a_0 - b_0 c_j}{\Delta_0} q_2 \prod_{k=1}^m \exp\left( +
\tilde{Y}_k \right) \Biggr].
   \label{eq:p1p1:mirror1}
\end{eqnarray}
This precisely matches the superpotential~(\ref{eq:pnpm:first}) derived
from integrating out fields in a different order, as expected.

As in the first ordering, there is a subtlety we have glossed over,
a multiplicative factor arising when integrating out some of the fields.
Here, the factor arises when integrating out $F_0$, $\tilde{F}_0$,
for the same reasons as before:  schematically,
the B/2 model path integral measure 
contains a factor of the form
\begin{equation}
\int d F_0 d \tilde{F}_0 \, \delta( a_0 F_0 + b_0 \tilde{F}_0 + \cdots)
\, \delta( c_0 F_0 + d_0 \tilde{F}_0 + \cdots),
\end{equation}
which again generates a numerical factor\footnote{
Tracing through this a bit more carefully, the numerical factor arises
from the delta functions, which arose from bosonic fields ($\sigma$'s), 
hence the
numerical factor is $\delta_0^{-1}$ instead of $( \Delta_0^{-1} )^{-1} = 
\Delta_0$ as one might have expected from a fermionic integral.
} of $\Delta_0^{-1}$ that
multiplies correlation functions, and which will be important 
in subsection~\ref{sect:pnpm:corrfns:ir}.

\subsection{Correlation functions in the lower-energy theory}
\label{sect:pnpm:corrfns:ir}

Next, we compute correlation functions in the new theory, 
for the case of ${\mathbb P}^1 \times {\mathbb P}^1$, obtained
after integrating out fields, and compare to the results for 
correlation functions computed in the UV theory, before integrating out
fields.  We will see an important subtlety.

Using the mirror (0,2) superpotential~(\ref{eq:p1p1:mirror1}),
and the operator mirror map
\begin{eqnarray}
\sigma & = & \frac{1}{\Delta_0} \left( d_0 \exp\left( - Y_0 \right)
\: - \: b_0 \exp\left( - \tilde{Y}_0 \right) \right),
\\
& = & \frac{1}{\Delta_0} \left( d_0 q_1 \exp\left( + Y_1 \right)
\: - \: b_0 q_2 \exp\left( + \tilde{Y}_1 \right) \right),
\\
\tilde{\sigma} & = & \frac{1}{\Delta_0} \left( a_0 \exp\left( - \tilde{Y}_0
\right) \: - \: c_0 \exp\left( - Y_0 \right) \right),
\\
& = & \frac{1}{\Delta_0} \left( a_0 q_2 \exp\left( + \tilde{Y}_1 \right)
\: - \: c_0 q_1 \exp\left( + Y_1 \right) \right),
\end{eqnarray}
where
\begin{equation}
\Delta_0 \: = \: a_0 d_0 \: - \: b_0 c_0,
\end{equation}
using the methods of \cite{Melnikov:2007xi},
we find that 
the two-point functions computed from the mirror above
are all $\Delta_0$ times the A/2 model correlation functions in
\cite{Closset:2015ohf}[section 4.2], reviewed in 
section~\ref{sect:p1p1:2pt:uv},
or in
other words,
\begin{equation}
\langle \sigma \sigma \rangle_{\rm mirror} \: = \: - \Delta_0
\frac{\Gamma_1}{\alpha},
\: \: \:
\langle \sigma \tilde{\sigma} \rangle_{\rm mirror} \: = \: + \Delta_0 
\frac{\Delta}{\alpha},
\: \: \:
\langle \tilde{\sigma} \tilde{\sigma} \rangle_{\rm mirror} \: = \: - \Delta_0
\frac{\Gamma_2}{
\alpha},
\end{equation}

However, we still need to take into account the subtlety discussed
in subsection~\ref{sect:pnpm:IR}.  Specifically, when deriving the (0,2)
Landau-Ginzburg model above from the UV presentation, we had to
perform changes-of-variables when integrating out fields, with the
effect that low-energy correlation functions should be multiplied by
factors of $1/\Delta_0$.  Taking that subtlety into account,
and dividing out the extra $\Delta_0$ factors, we find that the 
correct two-point functions precisely match 
both those of the A/2 model
\cite{Closset:2015ohf}[section 4.2], as well as those of the
original (UV) theory described in subsection~\ref{sect:pnpm:UV}.

It is also straightforward to compute four-point functions.
Their values in the A/2 model are given in
\cite{Chen:2016tdd}[appendix A.1].
When one computes them in the (lower-energy) Landau-Ginzburg model above,
not taking into account the subtlety discussed above,
one finds that the Landau-Ginzburg correlation functions are $\Delta_0$
times the A/2 model correlation functions.  Taking into account the
subtlety above, the overall factor of $1/\Delta_0$ multiplying all
correlation functions, fixes the four-point functions also.
In any event, once one knows that the two-point functions and the quantum
sheaf cohomology relations match, all of the higher-point functions are
guaranteed to match.

\subsection{Comparison to other (0,2) mirrors}
\label{sect:pnpm:compare}

Now, let us compare to the (0,2) mirrors in \cite{Chen:2016tdd,Gu:2017nye},
for brevity just for the case of ${\mathbb P}^1 \times
{\mathbb P}^1$.  As a matter of principle, these mirrors need not
necessarily match -- there could be multiple different UV theories
describing the same IR physics.  Nevertheless, in special
families, we will see that there is a match.

For example, in \cite{Chen:2016tdd}[section 4.2], it was argued that one (0,2) 
Landau-Ginzburg model those B/2 correlation functions correctly match
those of the corresponding A/2 theory on ${\mathbb P}^1 \times
{\mathbb P}^1$ had superpotential
\begin{equation}
W \: = \: F_1 J_1 \: + \: \tilde{F}_1 \tilde{J}_1,
\end{equation}
where
\begin{eqnarray}
J_1 & = & a X_1 \: - \: \frac{q_1}{X_1} \: + \: 
b \frac{ \tilde{X}_1^2}{ X_1 } \: + \: \mu \tilde{X}_1,
\\
\tilde{J}_1 & = & d \tilde{X}_1 \: - \: \frac{q_2}{\tilde{X}_1} \: + \:
c \frac{ X_1^2 }{\tilde{X}_1} \: + \: \nu X_1,
\end{eqnarray}
with
\begin{equation}
\mu \: = \: \det(A+B) - \det A - \det B,
\: \: \:
\nu \: = \: \det(C+D) - \det C - \det D,
\end{equation}
and operator mirror map
\begin{equation}
\sigma \: = \: X_1,
\: \: \:
\tilde{\sigma} \: = \: \tilde{X}_1.
\end{equation}

These expressions have the good property that they are in terms of
determinants of the matrices $A$, $B$, $C$, $D$, and so respect global
symmetries of the original theory.  For that matter,
the A/2 correlation functions only depend
upon those determinants, which is explicit in the mirrors constructed
in \cite{Chen:2016tdd}.

For purposes of comparison, for ${\mathbb P}^1 \times {\mathbb P}^1$,
the superpotential~(\ref{eq:p1p1:mirror1}) takes the form
\begin{eqnarray}
W & = &
- F_1 \left[
\exp\left( - Y_1 \right) \: - \: q_1 \frac{ ( a_1 d_0 - b_1 c_0 ) }{\Delta_0}
\exp\left( + Y_1\right) \: - \: q_2 \frac{ ( b_1 a_0 - a_1 b_0)}{\Delta_0}
\exp\left( + \tilde{Y}_1\right) 
\right]
\nonumber \\
& & 
\: - \:
\tilde{F}_1 \left[
\exp\left( - \tilde{Y}_1 \right) \: - \: q_1 \frac{ ( c_1 d_0 - d_1 c_0)}{
\Delta_0} \exp\left( + Y_1 \right) \: - \:
q_2 \frac{ ( d_1 a_0 - c_1 b_0 ) }{\Delta_0} \exp\left( + \tilde{Y}_1 \right)
\right].
\nonumber \\
& &
\end{eqnarray}
On the face of it, this
clearly does not match the mirror proposal of \cite{Chen:2016tdd}, and in fact,
is not even written in terms of global-symmetry-invariant determinants of
$A$, $B$, $C$, $D$.  Nevertheless, as we have seen, it does reproduce the
same correlation functions.

One could imagine using global symmetry transformations to rotate to
$a_0 = d_0=1, b_0=c_0 = 0$, the case considered in
\cite{Gu:2017nye}[section 5.1], in which case the result above reduces to
\begin{eqnarray}
W & = & - F_1 \left[
\exp\left( - Y_1\right) \: - \: q_1 a_1 \exp\left( + Y_1 \right)
\: - \: q_2 b_1 \exp\left( + \tilde{Y}_1 \right)
\right]
\nonumber \\
& & 
\: - \: \tilde{F}_1 \left[
\exp\left( - \tilde{Y}_1\right) \: - \: q_1 c_1 \exp\left( + Y_1 \right)
\: - \: q_2 d_1 \exp\left( + \tilde{Y}_1 \right)
\right].
\end{eqnarray}
In this case, 
\begin{equation}
a \: = \: a_1, \: \: \:
b \: = \: 0 \: = \: c, \: \: \:
d \: = \: d_1, \: \: \: 
\mu \: = \: b_1, \: \: \:
\nu \: = \: c_1,
\end{equation}
with operator mirror map
\begin{equation}
\sigma \: = \: q_1 \exp\left( + Y_1 \right),
\: \: \:
\tilde{\sigma} \: = \: q_2 \exp\left( + \tilde{Y}_1 \right).
\end{equation}
If we change variables as
\begin{equation}
\exp\left( - Y_0 \right) \: = \: q_1 \exp\left( + Y_1 \right),
\: \: \:
\exp\left( - \tilde{Y}_0 \right) \: = \: q_2 \exp\left( + 
\tilde{Y}_1 \right),
\end{equation}
then we can rewrite the superpotential as
\begin{eqnarray}
W & = & - F_1 \left[
q_1 \exp\left( + Y_0 \right) \: - \: a_1 \exp\left( - Y_0 \right)
\: - \: b_1 \exp\left( - \tilde{Y}_0 \right)
\right]
\nonumber \\
& & 
\: - \: \tilde{F}_1 \left[
q_2 \exp\left( + \tilde{Y}_0 \right) \: - \:
c_1 \exp\left( - Y_0 \right) \: - \:
d_1 \exp\left( - \tilde{Y}_0 \right)
\right],
\end{eqnarray}
which precisely matches the (0,2) mirror in \cite{Chen:2016tdd} for
the case $a_0 = d_0 = 1$, $b_0 = c_0 = 0$.  We will return to this case,
which also arose in \cite{Gu:2017nye}, in a more systematic analysis in
section~\ref{sect:previous}.

\section{Example:  Hirzebruch surfaces}
\label{sect:hirzebruch}

In this section we will compare to proposals for (0,2) mirrors to
Hirzebruch surfaces with a deformation of the tangent bunde,
as discussed in \cite{Chen:2017mxp}.  
Our analysis will follow the same form as that for
the mirror to ${\mathbb P}^n \times {\mathbb P}^m$, so we will be
comparatively brief.

A Hirzebruch surface ${\mathbb F}_n$
can be described by a GLSM with gauge group $U(1)^2$
and matter fields
\begin{center}
\begin{tabular}{c|cccc}
& $x_0$ & $x_1$ & $w$ & $s$ \\ \hline
$U(1)_1$ & $1$ & $1$ & $n$ & $0$ \\
$U(1)_2$ & $0$ & $0$ & $1$ & $1$
\end{tabular}
\end{center}
A deformation ${\cal E}$ of the tangent bundle is described
mathematically as the cokernel
\begin{equation}
0 \: \longrightarrow \: {\cal O}^2 \: \stackrel{*}{\longrightarrow} \:
{\cal O}(1,0)^2 \oplus {\cal O}(n,1) \oplus {\cal O}(0,1) \:
\longrightarrow \: {\cal E} \: \longrightarrow \: 0,
\end{equation}
where
\begin{equation}
* \: = \: \left[ \begin{array}{cc}
A x & Bx \\
\gamma_1 w & \beta_1 w \\
\gamma_2 s & \beta_2 s
\end{array} \right],
\end{equation}
and $x = [x_0, x_1]^T$.  In principle, additional nonlinear deformations
are also possible, but as they do not contribute to quantum sheaf cohomology
rings (see section~\ref{sect:proposal}), we omit them here.
The (2,2) locus corresponds to the case $A = I$, $B = 0$, $\gamma_1 = n$,
$\beta_1 = 1$, $\gamma_2=0$, $\beta_2=1$.

For a general (0,2) theory (with linear diagonal deformations), 
the $E$'s take the form
\begin{equation}
\overline{D}_+ \Lambda_{x, i} \: = \: ((\sigma A + \tilde{\sigma} B) x)_i,
\: \: \:
\overline{D}_+ \Lambda_w \: = \: (\gamma_1 \sigma + \beta_1 \tilde{\sigma}) w,
\: \: \:
\overline{D}_+ \Lambda_s \: (\gamma_2 \sigma + \beta_2 \tilde{\sigma}) s,
\end{equation}
where the $\Lambda$'s are the Fermi superfield partners to
the bosonic chiral fields.  Our mirror construction applies to diagonal
deformations, so we only consider the case that
\begin{equation}
\begin{array}{cc}
\overline{D}_+ \Lambda_{x,0} \: = \: ( a_0 \sigma + b_0 \tilde{\sigma}) x_0,
& 
\overline{D}_+ \Lambda_{x,1} \: = \: ( a_1 \sigma + b_1 \tilde{\sigma}) x_1,
\\ 
\overline{D}_+ \Lambda_w \: = \: (\gamma_1 \sigma + \beta_1 \tilde{\sigma}) w,
& 
\overline{D}_+ \Lambda_s \: = \: (\gamma_2 \sigma + \beta_2 \tilde{\sigma}) s.
\end{array}
\end{equation}

From our ansatz, the mirror Landau-Ginzburg model has fields
\begin{itemize}
\item $\sigma, \tilde{\sigma}$,
\item $(Y_{0,1}, F_{0,1})$, corresponding to $(x_{0,1}, \Lambda_{x,0-1})$
of the A/2 model,
\item $(Y_w, F_w)$, corresponding to $(w, \Lambda_w)$ of the A/2 model,
\item $(Y_s, F_s)$, corresponding to $(s, \Lambda_s)$ of the A/2 model,
\end{itemize}
and superpotential
\begin{eqnarray}
W & = &
\Upsilon_1 \left( Y_0 + Y_1 + n Y_w - t_1 \right) \: + \:
\Upsilon_2 \left( Y_w + Y_s - t_2 \right)
\nonumber \\
& &
\: + \: F_0 \left( a_0 \sigma + b_0 \tilde{\sigma} - \exp\left( - Y_0 \right)
\right)
\: + \: F_1 \left( a_1 \sigma + b_1 \tilde{\sigma} - \exp\left( - Y_1 \right)
\right)
\nonumber \\
& & 
\: + \: F_w \left( \gamma_1 \sigma + \beta_1 \tilde{\sigma} - 
\exp\left( - Y_w \right) \right)
\: + \: F_s \left( \gamma_2\sigma + \beta_2 \tilde{\sigma} - 
\exp\left( - Y_s \right) \right).
\end{eqnarray}

The operator mirror map is defined by the constraints imposed by the $F$'s:
\begin{eqnarray}
\exp\left( - Y_0 \right) & = & a_0 \sigma + b_0 \tilde{\sigma},
\\
\exp\left( - Y_1 \right) & = & a_1 \sigma + b_1 \tilde{\sigma},
\\
\exp\left( - Y_w \right) & = & \gamma_1 \sigma + \beta_1 \tilde{\sigma},
\\
\exp\left( - Y_s \right) & = & \gamma_2 \sigma + \beta_2 \tilde{\sigma},
\end{eqnarray}
and using the mirror D-term relations imposed by the $\Upsilon$'s, namely
\begin{equation}
\exp\left( - Y_0 - Y_1 - n Y_w \right) \: = \: q_1,
\: \: \:
\exp\left( - Y_w - Y_s \right) \: = \: q_2,
\end{equation}
we quickly derive the quantum sheaf cohomology (chiral ring) relations
\begin{equation}
\left( a_0 \sigma + b_0 \tilde{\sigma} \right)
\left( a_1 \sigma + b_1 \tilde{\sigma} \right)
\left( \gamma_1 \sigma + \beta_1 \tilde{\sigma} \right)^n \: = \: q_1,
\: \: \:
\left(  \gamma_1 \sigma + \beta_1 \tilde{\sigma} \right)
\left( \gamma_2 \sigma + \beta_2 \tilde{\sigma} \right) \: = \: q_2,
\end{equation}
or equivalently
\begin{equation}
\det \left( A \sigma + B \tilde{\sigma} \right)
\left( \gamma_1 \sigma + \beta_1 \tilde{\sigma} \right)^n \: = \: q_1,
\: \: \:
\left(  \gamma_1 \sigma + \beta_1 \tilde{\sigma} \right)
\left( \gamma_2 \sigma + \beta_2 \tilde{\sigma} \right) \: = \: q_2,
\label{eq:fn:qsc}
\end{equation}
which precisely match the known quantum sheaf cohomology ring relations
for this case \cite{McOrist:2007kp,McOrist:2008ji,Donagi:2011uz,Donagi:2011va}.

Next, we integrate out some of the fields to find a lower-energy effective
Landau-Ginzburg description of the same physics.  If we integrate out
$F_0$, $F_w$, we get the constraints
\begin{eqnarray}
a_0 \sigma \: + \: b_0 \tilde{\sigma} & = & \exp\left( - Y_0 \right),
\\
\gamma_1 \sigma \: + \: \beta_1 \tilde{\sigma} & = & \exp\left( - Y_w \right),
\end{eqnarray}
which can be solved to give
\begin{eqnarray}
\sigma & = & \frac{1}{\Delta_0} \left( \beta_1 \exp\left( - Y_0 \right)
\: - \: b_0 \exp\left( - Y_w \right) \right),
\label{eq:fn:opmirror1}
\\
\tilde{\sigma} & = & \frac{1}{\Delta_0} \left( a_0 \exp\left( - Y_w \right)
\: - \: \gamma_1 \exp\left( - Y_0 \right) \right),
\label{eq:fn:opmirror2}
\end{eqnarray}
for
\begin{equation}
\Delta_0 \: = \: a_0 \beta_1 - b_0 \gamma_1.
\end{equation}
Using the $\Upsilon$ constraints to eliminate $Y_0$, $Y_w$, we have
\begin{eqnarray}
\exp\left( - Y_w \right) & = & q_2 \exp\left( + Y_s \right),
\\
\exp\left( - Y_0 \right) & = & q_1 \exp\left( + Y_1 \right)
\exp\left( + n Y_w \right) \: = \:
(q_1 q_2^{-n} ) \exp\left( + Y_1 \right) \exp\left( - n Y_s \right),
\end{eqnarray}
and finally plugging in we get the lower-energy effective superpotential
\begin{eqnarray}
W & = &
F_1 \left( a_1 \sigma + b_1 \tilde{\sigma} - \exp\left( - Y_1 \right) \right)
\: + \:
F_s \left( \gamma_2 \sigma + \beta_2 \tilde{\sigma} - \exp\left( - Y_s \right)
\right),
\\
& = &
F_1  \left( \frac{ (a_1 \beta_1 - b_1 \gamma_1 ) }{\Delta_0} \exp\left( - Y_0 
\right)
\: + \:
\frac{ (-a_1 b_0 + b_1 a_0 ) }{\Delta_0} \exp\left( - Y_w \right)
\: - \: \exp\left( - Y_1 \right) \right)
\nonumber \\
& & \: + \:
F_s \left( \frac{ ( \gamma_2 \beta_1 - \beta_2 \gamma_1)}{\Delta_0}
\exp\left( - Y_0 \right) \: + \:
\frac{ ( - \gamma_2 b_0 + \beta_2 a_0 ) }{\Delta_0} \exp\left( - Y_w \right)
\: - \: \exp\left( - Y_s \right) \right),
\nonumber \\
& = & F_1 \biggl[  \frac{ (a_1 \beta_1 - b_1 \gamma_1 ) }{\Delta_0}
(q_1 q_2^{-n} ) \exp\left( + Y_1 \right) \exp\left( - n Y_s \right)
\: + \:
\frac{ (-a_1 b_0 + b_1 a_0 ) }{\Delta_0} q_2 \exp\left( + Y_s \right)
\nonumber \\
& & \hspace*{1.5in}
\: - \: \exp\left( - Y_1 \right) \biggr]
\nonumber \\
& &
\: + \: F_s \biggl[  \frac{ ( \gamma_2 \beta_1 - \beta_2 \gamma_1)}{\Delta_0}
(q_1 q_2^{-n} ) \exp\left( + Y_1 \right) \exp\left( - n Y_s \right)
\: + \:
\frac{ ( - \gamma_2 b_0 + \beta_2 a_0 ) }{\Delta_0}
q_2 \exp\left( + Y_s \right)
\nonumber \\
& & \hspace*{1.5in}
\: - \: \exp\left( - Y_s \right) \biggr].
\label{eq:fn:mirror:ir}
\end{eqnarray}
To be clear, because of the change of variables we performed in constraints
above, to match A/2 correlation functions, correlation functions in
this model must be multiplied by a factor of $1/\Delta_0$,
just as in our analysis in subsection~\ref{sect:pnpm:IR}.

As a consistency check, let us quickly verify from the
mirror~(\ref{eq:fn:mirror:ir}) above, plus the operator
mirror map~(\ref{eq:fn:opmirror1}), (\ref{eq:fn:opmirror2}), 
that the quantum sheaf cohomology
relations are obeyed.
Briefly,
\begin{eqnarray}
a_0 \sigma + b_0 \tilde{\sigma} & = & (q_1 q_2^{-n}) \exp\left( + Y_1 \right)
\exp\left( - n Y_s \right) \: \: \: \mbox{ from the operator mirror map},
\\
a_1 \sigma + b_1 \tilde{\sigma} & = & \exp\left( - Y_1 \right)
\: \: \: \mbox{ from the $F_1$ constraint},
\\
\gamma_1 \sigma + \beta_1 \tilde{\sigma} & = &
q_2 \exp\left( + Y_s \right) \: \: \:
\mbox{ from the operator mirror map},
\\
\gamma_2 \sigma + \beta_2 \tilde{\sigma} & = &
\exp\left( - Y_s \right) 
\: \: \:
\mbox{ from the $F_s$ constraint},
\end{eqnarray}
hence
\begin{eqnarray}
\left(a_0 \sigma + b_0 \tilde{\sigma} \right)
\left( a_1 \sigma + b_1 \tilde{\sigma} \right)
\left( \gamma_1 \sigma + \beta_1 \tilde{\sigma} \right)^n
& = & q_1,
\\
\left( \gamma_1 \sigma + \beta_1 \tilde{\sigma} \right)
\left( \gamma_2 \sigma + \beta_2 \tilde{\sigma} \right)
& = & q_2,
\end{eqnarray}
which are precisely the quantum sheaf cohomology ring 
relations~(\ref{eq:fn:qsc}) 
for this case.

Now, consider the mirror in the special case that 
$a_0 = 1$, $b_0 = 0$, $\beta_1 = 1$, $\gamma_1 = n$, in other
words, that they take their values on the (2,2) locus.
In this case, $\Delta_0 = 1$, and the mirror above becomes
\begin{eqnarray}
W & = & F_1 \left[ \left(a_1 - n b_1\right) (q_1 q_2^{-n}) \exp\left( + Y_1 \right)
\exp\left( - n Y_s \right)
\: + \:
b_1 q_2 \exp\left( + Y_s \right) \: - \:
\exp\left( - Y_1 \right) \right]
\nonumber \\
& &
\: + \: F_s \left[
\left( \gamma_2 - n \beta_2 \right) (q_1 q_2^{-n} ) \exp\left( + Y_1 \right)
\exp\left( - n Y_s \right)
\: + \:
\beta_2 q_2 \exp\left( + Y_s \right) \: - \:
\exp\left( - Y_s \right)
\right].  
\nonumber \\
\end{eqnarray}
Using the operator mirror map, we can write this more simply as
\begin{eqnarray}
W & = & F_1 \left[ a_1 \sigma \: + \: b_1 \tilde{\sigma}
\: - \: \exp\left( - Y_1 \right) \right]
\nonumber \\
& & \: + \: F_s \left[
\gamma_2 \sigma \: + \: \beta_2 \tilde{\sigma} \: - \: \exp\left(
- Y_s \right) \right].
\end{eqnarray}
Now, we can perform a change of variables to relate this to the
${\mathbb F}_n$ mirror described in \cite{Chen:2017mxp}[section 4.2],
\cite{Gu:2017nye}[section 5.2.1].  To relate to their notation,
if we define $X_1$, $X_3$ by
\begin{eqnarray}
\sigma & = & X_1 \: = \: \exp\left( - Y_0 \right),
\\
\tilde{\sigma} & = & X_3 - n X_1 \: = \:
\exp\left( - Y_w \right),
\end{eqnarray}
then the (0,2) superpotential above becomes
\begin{eqnarray}
W & = & F_1 \left[ a_1 X_1 \: + \: b_1 \left( X_3 - n X_1 \right)
\: - \: \frac{ q_1 }{ X_1 X_3^n } \right]
\nonumber \\
& &
\: + \: F_s \left[ \gamma_2 X_1 \: + \: \beta_2 \left( X_3 - n X_1 \right)
\: - \: \frac{q_2}{X_3} \right].
\end{eqnarray}
For the case we are considering ($a_0 = 1$, $b_0 = 0$, $\gamma_1 = n$,
$\beta_1 = 1$),
\begin{eqnarray}
a & = & \det A \: = \: a_1,
\\
b & = & \det B \: = \: 0,
\\
\mu_{AB} & = & b_1,
\end{eqnarray}
the coefficient of $F_1$ can be identified with the $J_1$ in 
\cite{Chen:2017mxp}[section 4.2], \cite{Gu:2017nye}[section 5.2.1],
and their $J_2$ is $n J_1$ plus the coefficient of $F_s$.
After a trivial linear rotation of $F_1$, $F_s$, we see that this
change of variables identifies, in this case, the (0,2) mirror
superpotential to ${\mathbb F}_n$ above, derived from our general
ansatz, with that discussed in \cite{Chen:2017mxp,Gu:2017nye}.
This matching was not necessary -- there can be different UV representations
of the same IR physics -- but it is certainly satisfying.
We will discuss a more general form of this construction in
section~\ref{sect:previous}.

\section{Example:  Grassmannians}
\label{sect:grassmannian}

So far all of our examples have involved abelian GLSMs.  We next turn
to a nonabelian example.
The Grassmannian $G(k,N)$ is
described by a $U(k)$ GLSM with chirals $\Phi^a_i$ and Fermis $\Psi^a_i$ 
in N copies of the fundamental representation, $a \in \{1,\cdots,k\}$, 
$i \in \{1,\cdots,N\}$. 
For linear and diagonal (0,2) deformations off the (2,2) locus
  \cite{Guo:2015caf}
\begin{equation}
\overline{D}_+ \Psi^a_i \: = \:
\left(\sigma^a_b + B_i^j \left( {\rm Tr} \, \sigma \right) \right) \Phi^b_j,
\end{equation}
where $B$ is diagonal,
$B={\rm diag}(b_1,\cdots,b_N)$.
The mirror theory consists of chiral fields $\sigma_a, Y_{ia}, X_{\mu\nu}$ 
and Fermi fields $\Upsilon_a, F_{ia}, \Lambda_{\mu\nu}$ with 
$a,\mu,\nu=1,\cdots,k, i=1,\cdots,N$ and $\mu \neq \nu$, 
in the notation of \cite{Gu:2018fpm}.  
For the fundamental representation of $U(k)$, the $a$-th component of the weight associated with $Y_{ib}$ is
\begin{equation}
\rho^a_{ib} = \delta^a_b
\end{equation}
and the roots are given by
\begin{equation}
\alpha_{\mu\nu}^a = \delta_\nu^a-\delta_\mu^a,
\end{equation}
therefore the superpotential reads
\begin{equation}
\begin{split}
W=&\sum_{a=1}^k \Upsilon_a \left( \sum_{i=1}^N Y_{ia} + \sum_{\mu \neq a}(\ln X_{a\mu} - \ln X_{\mu a}) - t \right)
\\
 &+ \: \sum_{i=1}^N \sum_{a=1}^k F_{ia} \left( \sigma_a + b_i \left(\sum_b \sigma_b\right) - \exp(-Y_{ia}) \right)
\: + \:
 \sum_{\mu \neq \nu} \Lambda_{\mu\nu} \left( 1+\frac{\sigma_\mu - \sigma_\nu}{X_{\mu\nu}} \right),
\end{split}  \label{eq:gkn:genlw}
\end{equation}
which gives the operator mirror map
\begin{equation}
\exp(-Y_{ia}) \: = \: \sigma_a + b_i \left(\sum_b \sigma_b \right).
\end{equation}

Next, we compute the excluded locus.
From the $X_{\mu \nu}$ poles,
since $X_{\mu \nu} = \sigma_{\nu} - \sigma_{\nu}$ along the critical locus,
we have
\begin{equation}
\sigma_a \: \neq \: \sigma_b
\end{equation}
for $a \neq b$.  That part is the same as on the (2,2) locus.
From the fact that $\exp(-Y) \neq 0$, the $F_{ia}$
coefficients imply that
\begin{equation}
\sigma_a \: + \: b_i \left( \sum_c \sigma_c \right) \: \neq \: 0,
\end{equation}
for all $a$ and $i$, which is a deformation of what one gets on the (2,2)
locus.

Let us take a moment to examine the second excluded locus condition further.
If we sum over $\sigma_a$, we get
\begin{equation}
\left( 1 + k b_i \right) \left( \sum_c \sigma_c \right) \: \neq \: 0
\end{equation}
for all $i$, hence for example 
\begin{equation}
1 + k b_i \: \neq \: 0
\end{equation}
for all $i$.  This condition is closely related to a constraint that arises
on the $B_i^j$ in order for the gauge bundle defined by the
$\overline{D}_+ \Psi$ to be a bundle, and not some more general sheaf.
Specifically, it was shown in \cite{Guo:2016suk}[theorem 3.3] that
the $B$'s define a bundle, and not a sheaf, if and only if there do not
exist $k$ eigenvalues of $B$ that sum to $-1$.  The excluded locus condition
we have just derived on the Coulomb branch implies that none of the $B$
eigenvalues equals $-1/k$, which is closely related.

Next, let us recover the A/2 model.
Upon integrating out $X_{\mu\nu}$ and $Y_{ia}$, we get
\begin{equation}
W_{\rm eff} \: = \: 
\sum_{a=1}^k \Upsilon_a \left( -\ln \prod_{i=1}^N \left(\sigma_a + b_i \left(\sum_b \sigma_b\right)\right) - t \right)
\end{equation}
and
\begin{equation}
H_X \: = \: \prod_{\mu \neq \nu} (\sigma_\mu - \sigma_\nu)^{-1},
\end{equation}
\begin{equation}
H_Y \: = \: \prod_{i=1}^N \prod_{a=1}^k \left(\sigma_a +
 b_i \left(\sum_b \sigma_b \right) \right),
\end{equation}
which reproduce the A/2 correlation functions of the $U(k)$ GLSM
\begin{equation}
\langle\mathcal{O}(\sigma)\rangle \: =  \:
\frac{1}{k!} \sum_{J^a_{\rm eff}=0} \frac{\mathcal{O}(\sigma)}{\left( \det_{a,b} \partial_b J^a_{\rm eff} \right) H_X H_Y}.
\end{equation}

Next, we shall integrate out some of the fields to construct a lower-energy
Landau-Ginzburg model in the pattern of \cite{Gu:2018fpm}[section 4.1].
Beginning with the (0,2) superpotential~(\ref{eq:gkn:genlw}),
integrating out the $\Upsilon_a$ gives the constraints
\begin{equation}
\sum_{i=1}^N Y_{ia} \: + \: \sum_{\mu \neq a} \ln \left( \frac{ X_{a \mu} }{
X_{\mu a} } \right) \: = \: t.
\end{equation}
Using these to eliminate $Y_{N a}$, we have
\begin{equation}
Y_{Na} \: = \: - \sum_{i=1}^{N-1} Y_{ia} \: - \:
\sum_{\mu \neq a} \ln \left( \frac{ X_{a \mu} }{ X_{\mu a} } \right)
\: = \: t,
\end{equation}
and so we define
\begin{eqnarray}
\Pi_a & = & \exp\left( - Y_{N a} \right), \\
& = & q \left[ \prod_{i=1}^{N-1} \exp\left( + Y_{i a} \right) \right]
\left[ \prod_{\mu \neq a} \frac{ X_{a \mu} }{ X_{\mu a} } \right],
\end{eqnarray}
which happens to match the $\Pi_a$ defined in the (2,2) mirror of $G(k,N)$
in \cite{Gu:2018fpm}[section 4.1].

Next, we integrate out $F_{N a}$, which gives constraints
\begin{equation}
\sigma_a \: + \: b_N \left( \sum_c \sigma_c \right) \: = \: 
\exp\left( - Y_{N a} \right) \: = \: \Pi_a.
\end{equation}
These equations can be solved to give
\begin{equation}
\sigma_a \: = \: \frac{1}{1 + k b_N} \left[ \left(1 + (k-1) b_N \right)
\Pi_a \: - \: b_N \sum_{c \neq a} \Pi_c \right].
\end{equation}
Plugging this back in, we get our expression for a mirror Landau-Ginzburg
theory:
\begin{eqnarray}
W & = &  \sum_{i=1}^{N-1} \sum_{a=1}^k F_{i a} \left(
\sigma_a + b_i \left( \sum_c \sigma_c \right) \: - \: 
\exp\left( - Y_{ia} \right) \right)
\nonumber \\
& & 
\: + \:
\sum_{\mu \neq \nu} \Lambda_{\mu \nu} \left( 1 \: + \:
\frac{ \sigma_{\mu} - \sigma_{\nu} }{ X_{\mu \nu} } \right),
\\
 & = & \sum_{i=1}^{N-1} \sum_{a=1}^k F_{i a} \left[
\frac{1}{1 + k b_N} \left( \left(1 + (k-1) b_N + b_i \right) \Pi_a \: + \:
 (b_i - b_N) \sum_{c \neq a} \Pi_c \right)
\: - \: \exp\left( - Y_{ia} \right) \right]
\nonumber \\
& &
\: + \:
\sum_{\mu \neq \nu} \Lambda_{\mu \nu} \left( 1 \: + \:
\frac{ \Pi_{\mu} - \Pi_{\nu} }{ X_{\mu \nu} }
\right).
\end{eqnarray}

As in earlier discussions, we have glossed over a subtlety:  when
integrating out the $F_{N a}$, we omitted a Jacobian factor of
\begin{equation}
\det({\rm Jac})^{-1} \: = \: \det\left[ \begin{array}{cccc}
1+b_N & b_N & \cdots & b_N \\
b_N & 1+b_N & \cdots & b_N \\
\vdots & & & \vdots \\
b_N & b_N & \cdots & 1+b_N \end{array} \right]^{-1}
\: = \: 
\frac{1}{1 + k b_N},
\end{equation}
which should be multiplied into correlation functions in order to match
against A/2 results.

As a consistency check, when all the $b_i = 0$, the (0,2) superpotential
above reduces to
\begin{eqnarray}
W & = &
\sum_{i=1}^{N-1} \sum_{a=1}^k F_{i a} \left( \Pi_a \: - \: \exp\left(
- Y_{ia} \right) \right)
\: + \:
\sum_{\mu \neq \nu} \Lambda_{\mu \nu} \left( 1 \: + \:
\frac{ \Pi_{\mu} - \Pi_{\nu} }{X_{\mu \nu} } \right),
\end{eqnarray}
which is precisely the (0,2) expansion of the (2,2) mirror superpotential
\begin{equation}
W \: = \: \sum_{i=1}^{N-1} \sum_{a=1}^k \exp\left( - Y_{ia} \right)
\: + \: \sum_{\mu \neq \nu} X_{\mu \nu} \: + \:
\sum_{a=1}^k \Pi_a
\end{equation}
computed in \cite{Gu:2018fpm}[section 4.1].

Next, we will derive the quantum sheaf cohomology relations from
this lower-energy Landau-Ginzburg model.
The $\Lambda_{\mu \nu}$ imply the constraints
\begin{equation}
X_{\mu \nu} \: = \: \Pi_{\nu} - \Pi_{\mu}
\end{equation}
along the critical locus, and similarly
from the $F_{ia}$,
\begin{eqnarray}
\exp\left( - Y_{ia} \right) & = &
\frac{1}{1 + k b_N} \left( \left(1 + (k-1) b_N + b_i \right) \Pi_a \: + \:
 (b_i - b_N) \sum_{c \neq a} \Pi_c \right),
\\
& = & \sigma_a \: + \: b_i \left( \sum_c \sigma_c \right)
\end{eqnarray}
along the critical locus.  Plugging into the definition of $\Pi_a$,
we have
\begin{equation}
\Pi_a \: = \: q \left[ \prod_{i=1}^{N-1} \exp\left( + Y_{ia} \right) 
\right] (-)^{k-1},
\end{equation}
hence
\begin{equation}
\Pi_a \prod_{i=1}^{N-1} \left[ \sigma_a \: + \:
b_i \left( \sum_c \sigma_c \right) \right] \: = \: (-)^{k-1} q,
\end{equation}
or more simply
\begin{equation}
\det\left( I \sigma_a + B ({\rm Tr}\, \sigma) \right) \: = \:
\prod_{i=1}^{N} \left[ \sigma_a \: + \:
b_i \left( \sum_c \sigma_c \right) \right] \: = \: (-)^{k-1} q,
\end{equation}
This is precisely the physical description of the quantum sheaf
cohomology ring relation in the A/2 model on $G(k,n)$ with the tangent bundle
deformation described above \cite{Guo:2015caf}, as expected.
Thus, we see this mirror correctly duplicates the quantum sheaf cohomology
ring.

Now, let us perform some consistency checks by computing correlation functions
in the mirror Landau-Ginzburg model above in two simple examples
and comparing to known results.

Our first example is the special case of $G(1,3) = {\mathbb P}^2$. This has no
mathematically 
nontrivial tangent bundle deformations, but nontrivial parameters can still
enter the GLSM and appear in correlation functions, 
and so it will give a nontrivial test.
In this case, the (0,2) superpotential above 
reduces to
\begin{eqnarray}
W & = & \sum_{i=1}^2 F_{i} \left[ \frac{1}{1+b_3} \left( 1 + b_i \right) \Pi
\: - \: \exp\left( - Y_{i} \right) \right],
\end{eqnarray}
with 
\begin{equation}
\Pi \: = \: q \prod_{i=1}^2 \exp\left( + Y_i \right),
\: \: \:
\sigma \: = \: \frac{1}{1 + b_3} \Pi.
\end{equation}
The matrix of derivatives of the superpotential terms is
\begin{equation}
(\partial_i J_j) \: = \: \frac{1}{1+b_3} \left[ \begin{array}{cc}
(1+b_1) \Pi \: + \: (1 + b_3) \exp\left( - Y_1 \right)
&
(1+b_1) \Pi \\
(1+b_2) \Pi &
(1+b_2) \Pi \: + \: (1 + b_3)\exp\left( - Y_2 \right)
\end{array} \right],
\end{equation}
and using the methods of \cite{Melnikov:2007xi}, we find
\begin{equation}
\langle \sigma^2 \rangle \: = \: \frac{1}{(1+b_1) (1+b_2) },
\: \: \:
\langle \sigma^5 \rangle \: = \: \frac{q}{(1+b_1)^2 (1+b_2)^2 (1+b_3) }.
\end{equation}
These are exactly $(1+b_3)$ times the A/2 correlation functions for this
model given in \cite{Guo:2015caf}[section 4.1], which are
\begin{equation}
\langle \sigma^2 \rangle \: = \: \frac{1}{(1+b_1) (1+b_2)(1+b_3) },
\: \: \:
\langle \sigma^5 \rangle \: = \: \frac{q}{(1+b_1)^2 (1+b_2)^2 (1+b_3)^2 }.
\end{equation}
As predicted, we multiply the (lower-energy) Landau-Ginzburg model correlation
functions by $1/(1+b_3)$ to get the A/2 model correlation functions.

Next, consider the case of $G(2,3) = {\mathbb P}^2$.  This model,
mirror to a $U(2)$ gauge theory,  again
has no mathematically nontrivial tangent bundle deformations, 
but will also serve as a 
test of correlation functions, as nontrivial parameters do enter the GLSM
and appear in correlation functions.  
Briefly, one now constructs a matrix of derivatives
of the functions multiplying $F_{11}$, $F_{12}$, $F_{21}$, $F_{22}$,
$\Lambda_{12}$, $\Lambda_{21}$, with respect to $Y_{11}$, $Y_{12}$,
$Y_{21}$, $Y_{22}$, $X_{12}$, $X_{21}$, and 
using the methods of \cite{Melnikov:2007xi}, we find
\begin{eqnarray}
\langle \sigma_1^2 \rangle & = & \frac{1 + 2 b_3}{\Delta} \left( -1 - 2 I_2
- 2 I_1 \right),
\\
\langle \sigma_1 \sigma_2 \rangle & = &
\frac{1 + 2 b_3}{\Delta} \left( 2 + 2 I_2 + 2 I_1 \right),
\\
\langle \sigma_2^2 \rangle & = & \frac{1 + 2 b_3}{\Delta} \left(
-1 - 2 I_2 - 2 I_1 \right),
\end{eqnarray}
where, following the notation of \cite{Guo:2015caf},
\begin{eqnarray}
I_1 & = & \sum_i b_i,
\\
I_2 & = & \sum_{i < j} b_i b_j,
\\
I_3 & = & b_1 b_2 b_3,
\\
\Delta & = & 2 \prod_{i < j} \left( 1 + b_i + b_j \right).
\end{eqnarray}
The correlation functions above are precisely $(1+2 b_3)$ times the
A/2 model correlation functions computed in \cite{Guo:2015caf},
precisely as expected from the normalization subtlety discussed
in section~\ref{sect:pnpm:IR}.

\section{Example:  Flag manifolds}
\label{sect:flag}

In this section, we will briefly outline mirrors to flag manifolds.
The GLSM describing the flag manifold $F(k_1,k_2,\cdots,k_n,N)$ is a quiver gauge theory with gauge group $U(k_1) \times \cdots \times U(k_n)$
\cite{Donagi:2007hi}. For each $s=1,\cdots,n-1$, there is a chiral multiplet
$\Phi_{s,s+1}$ and a Fermi multiplet $\Psi_{s,s+1}$ transforming in the fundamental representation of
$U(k_s)$ and in the antifundamental representation of $U(k_{s+1})$.
There are also chiral multiplets $\Phi_{n,n+1}^i$ and Fermi multiplets $\Psi_{n,n+1}^i$ transforming in
the fundamental representation of $U(k_n)$ for $i=1,\cdots,N$. The $E$-terms of this theory are given by \cite{Guo:2018iyr}
\begin{equation}
\begin{split}
&\overline{D}_+ \Psi_{s,s+1}
\: = \: \Phi_{s,s+1} \Sigma^{(s)} \: - \:
\Sigma^{(s+1)} \Phi_{s,s+1} \: + \: \sum_{t=1}^n u^s_t \left({\rm Tr}\, \Sigma^{(t)} \right)
\Phi_{s,s+1},
\\
&\quad s=1,\cdots,n-1,
\\
&\overline{D}_+ \Psi_{n,n+1}^i
\: = \:
 \Phi_{n,n+1} \Sigma^{(n)} \: + \: \sum_{t=1}^n \left({\rm
Tr}\, \Sigma^{(t)}\right) {A_t}_j^i \Phi_{n,n+1}^j, ~i,j=1,\cdots,N.
\end{split}
\end{equation}
The matrices $A_t$ are assumed to be diagonal in this paper, i.e.
\begin{equation}
{A_t}_j^i = A_{ti} \delta^i_j.
\end{equation}
The mirror theory is a Landau-Ginzburg model consisting of chiral fields
\begin{equation}
\sigma_{a_s}^{(s)},~ {Y^{(s)}}^{a_s}_{b_s},~ X^{(s)}_{\mu_s\nu_s}
\end{equation}
and Fermi fields
\begin{equation}
\Upsilon_{a_s}^{(s)},~ {F^{(s)}}^{a_s}_{b_s},~ \Lambda^{(s)}_{\mu_s\nu_s}
\end{equation}
for $s=1,\cdots,n$, $a_s=1,\cdots,k_s$, $b_s=1,\cdots,k_{s+1}$, $\mu_s,\nu_s=1,\cdots,k_s$ and $\mu_s \neq \nu_s$ where $k_{n+1}=N$.

The superpotential is
\begin{equation}
\begin{split}
W=&\sum_{s=1}^n\sum_{a_s=1}^{k_s} \Upsilon^{(s)}_{a_s} \left( \sum_{b_s=1}^{k_{s+1}} {Y^{(s)}}^{a_s}_{b_s} - \sum_{\alpha_s=1}^{k_{s-1}}{Y^{(s-1)}}^{\alpha_s}_{a_s} + \sum_{\mu_s \neq a_s}(\ln X^{(s)}_{a_s\mu_s} - \ln X^{(s)}_{\mu_s a_s}) - t_s \right)
\\
 &+ \: \sum_{s=1}^n \sum_{a_s=1}^{k_s} \sum_{b_s=1}^{k_{s+1}} {F^{(s)}}^{a_s}_{b_s} \left( {E^{(s)}}^{a_s}_{b_s}(\sigma) - \exp\left(-{Y^{(s)}}^{a_s}_{b_s}\right) \right) 
\\
&  + \:
\sum_{s=1}^n \sum_{\mu_s \neq \nu_s} \Lambda^{(s)}_{\mu_s\nu_s} \left( 1+\frac{\sigma^{(s)}_{\mu_s} - \sigma^{(s)}_{\nu_s}}{X^{(s)}_{\mu_s\nu_s}} \right),
\end{split}
\end{equation}
where $k_0=0$,
\begin{equation}
{E^{(s)}}^{a_s}_{b_s}(\sigma) \: = \:
 \sigma^{(s)}_{a_s} \: - \: \sigma^{(s+1)}_{b_s}
\: + \: \sum_{t=1}^n u^s_t {\rm Tr}\, \sigma^{(t)}
\end{equation}
for $s=1,\cdots,n$, $a_s=1,\cdots,k_s$, $b_s=1,\cdots,k_{s+1}$ and
\begin{equation}
{E^{(n)}}^{a_n}_{b_n}(\sigma) \: = \:
 \sigma^{(n)}_{a_n} \: + \:  \sum_{t=1}^n A_{tb_n} {\rm Tr} \, \sigma^{(t)}
\end{equation}
for $a_n=1,\cdots,k_n$ and $b_n=1,\cdots,N$. Again, integrating out $X^{(s)}_{\mu_s\nu_s}$ and $\Lambda^{(s)}_{\mu_s\nu_s}$ shifts the FI parameters
\begin{equation}
t_s \rightarrow t_s + (k_s-1) \pi i.
\end{equation}

\section{Hypersurfaces}
\label{sect:hypersurfaces}

So far, our examples have involved mirrors to GLSMs without a superpotential.
One can add a superpotential to the original theory, following the
same prescription as \cite{Gu:2018fpm}; namely, one assigns R-charges to the
fields, and then takes the mirrors to fields with nonzero R charges,
following the same pattern as in \cite{Gu:2018fpm}.
For example, if a chiral field $\phi$ of the original theory has
R-charge $r$, then the fundamental field in the mirror is
\begin{equation}
X \: \equiv \: \exp\left( - (r/2) Y \right),
\end{equation}
and the theory has a ${\mathbb Z}_{2/r}$ orbifold.

As a result, the mirror (0,2) theory does not depend upon the details
of the original superpotential, only upon R-charges.  For (2,2) theories,
such statements are standard, but in (0,2) theories, they have come to 
be believed only somewhat more recently \cite{McOrist:2008ji}, and only
as statements about GLSM descriptions.  In any event, the point is that
our mirror construction implicitly reproduces the conjecture of 
\cite{McOrist:2008ji} that A/2-twisted GLSMs are independent of 
precise superpotential terms, and depend only upon R-charges.

\section{Conclusions}

In this paper we have described an extension of the nonabelian 
mirror proposal of \cite{Gu:2018fpm} from two-dimensional (2,2) supersymmetric
theories to (0,2) supersymmetric theories.  The result is a simple systematic
ansatz which both generalizes and simplifies 
previous approaches to Hori-Vafa-style
(0,2) abelian mirrors \cite{Adams:2003zy,Chen:2016tdd,Chen:2017mxp,Gu:2017nye},
and also applies to nonabelian cases
\cite{Gu:2018fpm,Chen:2018wep,Gu:2019zkw}.  We have
demonstrated that this mirror proposal has the desired properties
of a gauge theoretic mirror:  it reproduces symmetries,
correlation functions
and quantum sheaf cohomology rings, and demonstrated how one can recover
the one-loop effective superpotential of the original theory, in general
cases.  In addition,
we
have checked the proposal in specific
examples of mirrors in abelian and nonabelian theories.

\section{Acknowledgements}

We would like to thank Z.~Chen and I.~Melnikov for useful discussions.
W.G. would like to thank the math department of Tsinghua University
for hospitality while this work was completed,
and
E.S. would like to thank the Aspen Center for Physics for hospitality while
this work was completed.  The Aspen Center for Physics is supported by
National Science Foundation grant PHY-1607611.
E.S. was partially supported by NSF grant PHY-1720321.

\end{document}